\documentclass[twoside]{article}

\def\FL{}    

  \makeatletter
\def\artsectnumbering{%
  \@addtoreset{equation}{section}   
  \def\theequation{\thesection.\arabic{equation}}}
  \makeatother

\artsectnumbering

\def\Eq{Eq.\ } \def\Eqs{Eqs.\ }   

\def\book#1{``#1,"}    \def\jour#1{{\it #1\/}}

\def\nn{\nonumber} 

\catcode`@=11     
     
\def\eqalign#1{\null\,\vcenter{\openup\jot\m@th
  \ialign{\strut\hfil$\displaystyle{##}$&$\displaystyle{{}##}$\hfil
      \crcr#1\crcr}}\,}

\def\meqalign#1{\null\,\vcenter{\openup\jot\m@th
  \ialign{\strut\hfil$\displaystyle{##}$&&$\displaystyle{{}##}$\hfil
      \crcr#1\crcr}}\,}

\def\imeqalign#1{\null\,\vcenter{\openup\jot\m@th
  \ialign{\strut\hfil$\displaystyle{##}$&$\displaystyle{{}##}$\hfil
         &&\hfil$\displaystyle{##}$&$\displaystyle{{}##}$\hfil
      \crcr#1\crcr}}\,}

\catcode`@=12   

\def\leftcontract{\mathop{\hbox{\vrule height0.5pt width6pt \vrule width0.5pt 
   height6pt}}}
\def\rightcontract{\mathop{\hbox{\vrule width0.5pt height6pt%
  \vrule height0.5pt width6pt}}}

\def\Lie{\hbox{\it\char'44}}       

\def\half{{\textstyle\frac{1}{2}}}      
\def\fraction#1#2{{\textstyle\frac{#1}{#2}}}   

\def\Tr{\mathop{\rm Tr}\nolimits}       

\def\grad{\mathop{\rm grad}\nolimits}  \def\del{\nabla}               
\def\div{\mathop{\rm div}\nolimits}    \def\curl{\mathop{\rm curl}\nolimits}

\def\SYM{\mathop{\rm SYM}\nolimits}    \def\ALT{\mathop{\rm ALT}\nolimits}
     
\def\oversymbol#1#2{\vbox{\ialign{##\crcr \hfil$#1$\hfil\crcr
   \noalign{\kern1pt\nointerlineskip}%
   \hbox{$\hfil\displaystyle#2\hfil$}\crcr}}}
\def\overeq#1{\oversymbol{\scriptstyle\kern.5pt =}{#1}}

  \def\sub#1{_{\hbox{$#1$}}}  

\def\four{{}^{(4)}\kern-1pt}
\def\dual{{}^\ast\kern-1.5pt}  
\def\dualp#1{{}^{\ast_{(\hbox{$\scriptstyle #1$})}} \kern-1pt}

\def\subrp#1{_{\rm(#1)}} \def\suprp#1{^{\rm(#1)}}
\def\lrs{\subrp{lrs}}
\def\refe{\subrp{ref}}
\def\refr{\subrp{refr}}
\def\ind{\subrp{ind}}

\def\thom{\subrp{thom}}
\def\geo{\subrp{geo}}
\def\gmag{\subrp{gm}}
\def\gyro{\subrp{gyro}}
\def\so{\subrp{so}}
\def\sc{\subrp{sc}}

\def\lief{\subrp{lie\flat}}
\def\lie{\subrp{lie}}
\def\tem{\subrp{tem}}
\def\fw{\subrp{fw}}

\def\cfw {\subrp{cfw}}

\def\sym {\subrp{sym}}
\def\perpp {\suprp{\bot}}
\def\parp {\suprp{||}}
\def\em {\suprp{EM}}
\def\g {\suprp{G}}
\def\TR {\suprp{TR}}

\def\subpmath#1{_{(#1)}}

\newcommand{\pubheading}
{\setbox43=\vbox{
\flushleft{
{\footnotesize Annals of Physics \hfill
          Vol.\ 215, No.\ 1, April 1992, pp.~1--50 }\\[-2pt]
	{\footnotesize \copyright 1992 by Academic Press 
          \hfill [Reformatted with corrections 2001]
}\\
}}
\setbox44=\vbox to 0pt{\vskip-22.5pt\vskip-25pt 
\hbox to \textwidth{\vbox to 8.5pt{}\box43}\vss}
\vspace*{-11.1cm}\box44\vspace*{11.1cm}\nointerlineskip} 

\input prepictex \input pictex \input postpictex
\def\mathput#1{\relax \ifmmode \displaystyle #1\else $\displaystyle #1$\fi}

\begin{document}

\title{The Many Faces of Gravitoelectromagnetism}

\author{Robert T. Jantzen
       \vadjust{\vskip3pt}  \\ \small\it
Department of Mathematical Sciences, Villanova University,\\ \small\it
Villanova, PA 19085, USA\\ \small\it
and International Center for Relativistic Astrophysics\\ \small\it
University of Rome, I-00185 Roma, Italy
       \vadjust{\vskip3pt} \\
Paolo Carini
       \vadjust{\vskip3pt} \\ \small\it
International Center for Relativistic Astrophysics\\ \small\it
University of Rome, I-00185 Roma, Italy\\ \small\it
and GP-B, Hansen Labs, Stanford University, Stanford, CA 94305, USA
      \vadjust{\vskip3pt} \\
and
      \vadjust{\vskip3pt} \\
Donato Bini
       \vadjust{\vskip3pt}\\ \small\it
International Center for Relativistic Astrophysics\\ \small\it
University of Rome, I-00185 Roma, Italy\\
\null\\ \small
Received August 23, 1991}

\date{}

\maketitle
\pubheading

\begin{abstract}

The numerous ways of introducing spatial gravitational forces are
fit together in a single framework enabling their interrelationships
to be clarified. 
This framework is then used to treat
the ``acceleration equals force" equation and
gyroscope precession, both of which are then discussed in
the post-Newtonian approximation, followed by 
a brief examination of the Einstein equations
themselves in that approximation.

\end{abstract}

\section{Introduction}

The concept of spatial gravitational forces modeled after the electromagnetic
Lorentz force has a long history and many names associated with it 
\cite{lanlif41}--\cite{damsofxua}.
Born in the Newtonian context of centrifugal and Coriolis forces introduced by
a rigidly rotating coordinate system in a flat Euclidean space, it has
found a number of closely related but
distinct generalizations within the context of general relativity
and its linearized approximation.
With the frequent reference to ``gravitoelectromagnetism" occurring in recent
literature, it is time to place all of these notions of ``noninertial
forces" into a single framework which in turn may be used to infer
relationships among them.

Key to all of these notions is the splitting of spacetime into ``space plus
time", accomplished locally by means of 
an observer congruence, namely a congruence
of timelike worldlines with (future-pointing) unit tangent vector field $u$
which may be interpreted as the 4-velocity field of a family of test observers
filling the spacetime or some open submanifold of it 
\cite{ehl}--\cite{sacwu}.
These worldlines
have a natural parametrization by the proper time $\tau_u$ measured along them
and defined to within an initial value on each worldline.
The orthogonal decomposition of each tangent space into a local time direction
along $u$ and the orthogonal local rest space $LRS_u$ may be used to decompose
all spacetime tensors and tensor equations into a ``space plus time"
representation, i.e., to ``measure" them.
This leads to a family of ``spatial" spacetime tensor
fields (giving zero upon any contraction with $u$)  which represent each
spacetime field and a family of spatial equations which 
represent each spacetime equation.

Such a splitting permits a better interface of our 3-dimensional intuition
and experience with the 4-dimensional spacetime geometry in certain
gravitational problems, though it may complicate others.
It can be particularly useful in spacetimes which have a geometrically
defined timelike congruence, either explicitly given or defined implicitly
as the congruence of orthogonal trajectories to a slicing (foliation)
of spacetime by a family of spacelike hypersurfaces, the latter leading to
a timelike congruence with vanishing vorticity or ``rotation" (the
hypersurface-forming condition for the distribution $LRS_u$).
Stationary spacetimes have a preferred congruence of Killing trajectories
associated with the stationary symmetry, which is timelike on an open
submanifold of spacetime.
Stationary axially symmetric spacetimes have in addition a preferred slicing
whose orthogonal trajectories coincide with the worldlines of
locally nonrotating test observers on an open submanifold of 
spacetime 
\cite{bar70}--\cite{greschvis}.
Cosmological spacetimes with a spatial homogeneity subgroup have a
preferred spacelike slicing by the orbits of this subgroup.

A partial splitting of spacetime based only on a timelike congruence
(splitting off time alone) or a spacelike slicing (splitting off space alone)
will be referred to as the congruence and hypersurface splittings respectively.
Often a congruence and transversal slicing occur in the same context,
with at least one of the components satisfying the causality condition
of the corresponding splitting. Such a pair will be said to define
a ``nonlinear reference frame" (to avoid confusion with existing terms)
and a full splitting of spacetime into
``time plus space" ($1+3$) or ``space plus time" ($3+1$) 
respectively 
\cite{lic}--\cite{ger}.
Introduce the suggestive term ``threading" parallel to the term ``slicing,"
in order to describe the transversal congruence which ``threads" 
(by transversality) the slicing. The two full splittings will be called
the threading splitting (timelike threading) and the slicing splitting
(spacelike slicing).  Each has an associated observer congruence of the
corresponding partial splitting. When both causality conditions hold, both
splittings are valid and one may transform between them, unless the
nonlinear reference frame is orthogonal (orthogonal slicing and threading),
in which case they coincide.
The nonlinear reference frame itself provides another splitting
which is often used to represent the former two, namely the
(in general)
nonorthogonal splitting of the tangent spaces into the local
threading direction and the local slicing directions.
This will be called the reference splitting.

In addition, the threading or slicing may be provided with a parametrization,
namely a class of affinely related parameters on each congruence curve
or of the family of slices respectively. In a parametrized nonlinear
reference frame, both components may be compatibly parametrized, with their 
parametrizations linked in an obvious way. In a stationary spacetime
the canonical parameter on the orbits of the stationary symmetry provides
a natural parametrization for the timelike
Killing threading, while in a spatially
homogeneous spacetime the proper time measured orthogonally to the family
of geodesically parallel spatial hypersurfaces of homogeneity is a natural 
parameter for that preferred slicing.

Spatial gravitational forces have been defined in all of these
contexts, depending on or independent of the parametrizations,
both in the fully nonlinear theory as well as in the linearized theory.
The proper question to ask is not which of these various
descriptions to choose is the ``best" or ``correct" one,
but what exactly each one of them measures and which is particularly
suited to a particular application where it can help provide
intuition about  or simplify the presentation of the invariant spacetime
geometry that all of them may be used to reconstruct.
Until now there has been no effort to clarify the interrelationships between
the many different approaches favored by numerous groups working with
isolated formalisms. A true relativity of formalism is needed to
break the barrier to a more versatile application of multiple
approaches whose selection is determined by the application and not by
the inertia of the investigator.  A careful development of this relativity
of formalism, as well as an appropriate historical survey of the topic,
requires a more lengthy exposition \cite{jancar91},
so only a brief sketch will be presented here, limiting historical
credit to references in the text.

The slicing point of view, often called the ADM approach \cite{arndesmis},
has been effectively promoted by the textbook by Misner, Thorne and
Wheeler \cite{misthowhe}, whose conventions will be assumed unless otherwise
indicated. The same effective notation and terminology will be extended
to the threading point of view, partially presented in the textbook by
Landau and Lifshitz \cite{lanlif75}.

\section{Observer-orthogonal splitting}

Let $\four{\rm g}$ (signature {\tt -+++} 
and components $\four g_{\alpha\beta}$,
$\alpha,\beta,\ldots =0,1,2,3$)
be the spacetime metric, $\four\del$ its associated covariant derivative
operator, and $\four\eta$ the unit volume 4-form 
which orients spacetime
($\four\eta_{0123} = \four g^{1/2}$ 
in an oriented frame, where $\four g \equiv |\det(\four g_{\alpha\beta})|$).
Assume the spacetime is also time oriented and 
let $u$ be a future-pointing unit timelike vector field 
($u^\alpha u_\alpha = -1$)
representing the 4-velocity field of a family of test observers filling
the spacetime (or some open submanifold of it).
If $S$ is an arbitrary tensor field, let $S^\flat$ and $S^\sharp$ denote
its totally covariant and totally contravariant forms with respect to the
metric index-shifting operations. It is also convenient to introduce the
right contraction notation $ [S\rightcontract X]^\alpha = S^\alpha{}_\beta
X^\beta$ for the contraction of a vector field and the covariant index of
a $1\choose1$-tensor field, representing the action of a linear transformation
of each tangent space into itself. 
In general let the
left contraction $S\leftcontract T$ denote the tensor product of the
two tensors $S$ and $T$ with a contraction between the rightmost contravariant
index of $S$ with the leftmost covariant index of $T$
(i.e., $ S^{\ldots\alpha}_{\ \ldots} T^{\ldots}_{\ \alpha\ldots} $),
and let the right
contraction $S\rightcontract T$ denote the tensor product with a contraction
between the leftmost contravariant index of $T$ with the rightmost
covariant index of $S$
(i.e., $ S^{\ldots}_{\ \ldots\alpha} T^{\alpha\ldots}_{\ \ldots} $),
assuming in each case that such indices exist.
For a ${1\choose1}$-tensor field $S$, let $S^2 \equiv S\rightcontract S$.

The observer-orthogonal decomposition of the tangent space, and in turn of
the algebra of spacetime tensor fields, is accomplished by the temporal
projection operator $T(u)$ along $u$ and the spatial projection operator
$P(u)$ onto $LRS_u$, which may be identified with mixed second rank
tensors acting by contraction
\begin{equation}\eqalign{
    \delta^\alpha{}_\beta &= T(u)^\alpha{}_\beta +  P(u)^\alpha{}_\beta \ ,\cr
   T(u)^\alpha{}_\beta &=  - u^\alpha u_\beta \ , \cr
   P(u)^\alpha{}_\beta &=  \delta^\alpha{}_\beta + u^\alpha u_\beta \ .\cr
}\end{equation}   
These satisfy the usual orthogonal projection relations  
$P(u)^2 = P(u)$,
$T(u)^2 = T(u)$, and $T(u)\rightcontract P(u) = P(u) \rightcontract T(u) = 0$.
Let 
\begin{equation}
      [ P(u) S ]^{\alpha\ldots}_{\ \ \beta\ldots} =
       P(u)^\alpha{}_\gamma \cdots P(u)^\delta{}_\beta \cdots
       S^{\gamma\ldots}_{\ \ \delta\ldots} 
\end{equation}   
denote the spatial projection of a tensor $S$ on all indices.  

The ``measurement of $S$" by the observer congruence is the family of spatial
tensor fields which result from the spatial projection of
all possible contractions of $S$ by any number of factors of $u$.
For example, if $S$ is a $1\choose1$-tensor, then its measurement
\begin{equation}\eqalign{
     S^\alpha{}_\beta \leftrightarrow &
     ( u^\delta u_\gamma S^\gamma{}_\delta,
         P(u)^\alpha{}_\gamma u^\delta S^\gamma{}_\delta, 
         P(u)^\delta{}_\alpha u_\gamma S^\gamma{}_\delta,
         P(u)^\alpha{}_\gamma P(u)^\delta{}_\beta S^\gamma{}_\delta ) \cr
}\end{equation}   
results in a scalar field, a spatial vector field, a spatial 1-form and
a spatial $1\choose1$-tensor field. It is exactly this family of fields
which occur in the (orthogonal) ``decomposition of $S$" with respect to
the observer congruence
\FL
\begin{equation}\eqalign{
    S^\alpha{}_\beta 
        &= [T(u)^\alpha{}_\gamma + P(u)^\alpha{}_\gamma ] 
         [T(u)^\delta{}_\beta + P(u)^\delta{}_\beta ] 
             S^\gamma{}_\delta \cr
    &= [u^\delta u_\gamma S^\gamma{}_\delta] u^\alpha u_\beta
       + \cdots + [ P(u) S ]^\alpha{}_\beta \ .\cr
}\end{equation}   
The spatial metric $[ P(u) \four{\rm g} ]_{\alpha\beta} = P(u)_{\alpha\beta}$
and the spatial unit volume 3-form 
$\eta(u)_{\alpha\beta\gamma} = u^\delta \four\eta_{\delta\alpha\beta\gamma}
= [P(u) \, u \leftcontract \four\eta]_{\alpha\beta\gamma}$
are the only nontrivial spatial fields which result from the measurement
of the spacetime metric and volume 4-form.

Introduce also the spatial Lie derivative \cite{isenes}
$ \Lie(u)\sub{X} = P(u)\Lie\sub{X}$ by the vector field $X$, the spatial
exterior derivative $ d(u) = P(u) d$, the spatial covariant derivative
$ \del(u) = P(u) \four \del$, the spatial Fermi-Walker derivative
(``Fermi-Walker temporal derivative") $\del\fw(u)= P(u) \four\del\sub{u}$
and the Lie temporal derivative
$\del\lie(u) = P(u) \Lie\sub{u} = \Lie(u)\sub{u}$.
Note that these spatial differential operators do not obey the usual
product rules for nonspatial fields since undifferentiated
factors of $u$ are killed by the spatial projection.

It is convenient to introduce 3-dimensional vector notation for the
spatial inner product and spatial cross product of two spatial vector fields
$X$ and $Y$. The inner product is just
\begin{equation}
      X \cdot_u Y  = P(u)_{\alpha\beta} X^\alpha Y^\beta 
\end{equation}   
while the cross product is
\begin{equation}
      [ X \times_u Y ]^\alpha = \eta(u)^\alpha{}_{\beta\gamma} X^\beta Y^\gamma
    \ .     
\end{equation}   
If one lets $\vec\del(u)$ be the ``vector derivative operator" 
$\del(u)^\alpha$,
then one can introduce spatial gradient, curl and divergence operators for
functions $f$ and spatial vector fields $X$ by
\begin{equation}\eqalign{
     \grad_u f &= \vec\del(u) f = [d(u) f]^\sharp \ ,\cr
     \curl_u X &= \vec\del(u) \times_u X 
                      = [\dualp{u} d(u) X^\flat ]^\sharp\ ,\cr
     \div_u  X &= \vec\del(u) \cdot_u X 
                     = \dualp{u} [ d(u) \dualp{u} X^\flat ] \ ,\cr
}\end{equation}   
where $\,\dualp{u}\,$ is the spatial duality operation for antisymmetric
tensor fields associated with the spatial volume form $\eta(u)$ in the usual
way. These definitions enable one to mimic all the usual formulas of 
3-dimensional vector analysis. The spatial exterior derivative formula
for the curl has the index form
\begin{equation}
   [ \curl_u X ]^\alpha = \eta(u)^{\alpha\beta\gamma} \four\del_\beta X_\gamma
\end{equation}   
and also defines a useful operator for nonspatial vector fields $X$.

Measurement of the covariant derivative $[\four\del u]^\alpha{}_\beta
= u^\alpha{}_{;\beta}$ leads to two spatial fields, the
acceleration vector field $a(u)$ and the kinematical mixed tensor field $k(u)$
\begin{equation}\eqalign{
  u^\alpha{}_{;\beta} &= -a(u)^\alpha u_\beta - k(u)^\alpha{}_\beta \ ,\cr
  a(u) &= \del\fw (u) u \ ,  \cr 
 k(u) &= - \del(u) u = \omega(u) - \theta(u) \ .\cr
}\end{equation}   
The kinematical tensor field may be decomposed into its antisymmetric and
symmetric parts 
\cite{ehl}--\cite{ell73,misthowhe,hawell}
\begin{equation}\eqalign{
  [\omega(u)^\flat]_{\alpha\beta} &=  
P(u)^\delta{}_\beta P(u)^\gamma{}_\alpha \typeout{important correction}
u_{[\delta;\gamma]} 
                  = \half [ d(u) u^\flat ]_{\alpha\beta} \ ,\cr
  [\theta(u)^\flat]_{\alpha\beta}  &= 
P(u)^\delta{}_\beta P(u)^\gamma{}_\alpha \typeout{important correction}
u_{(\beta;\alpha)} 
           =  \half [ \del\lie(u) P(u)^\flat ]_{\alpha\beta} 
           = \half \Lie(u)\sub{u} \four g_{\alpha\beta} \ ,\cr
}\end{equation}   
defining the mixed rotation or vorticity
tensor field $\omega(u)$ (whose sign depends on convention)
and the mixed expansion
tensor field $\theta(u)$, the latter of which may itself be decomposed into
its tracefree and pure trace parts
\begin{equation}
     \theta(u) = \sigma(u) + \frac13 \Theta(u) P(u) \ ,
\end{equation}   
where the mixed shear tensor field $\sigma(u)$ is tracefree 
($\sigma(u)^\alpha{}_\alpha = 0$)
and the expansion scalar is
\begin{equation}
          \Theta(u) = u^\alpha{}_{;\alpha} 
          = \dualp{u}[\del\lie(u) \eta(u)] \ .
\end{equation}   
Define also the rotation or vorticity vector field 
$\vec\omega(u)= \half \curl_u u$ as the spatial dual of the spatial rotation
tensor field
\begin{equation}
    \omega(u)^\alpha = 
       \half \eta(u)^{\alpha\beta\gamma} \omega(u)_{\beta\gamma}
      = \half \four\eta^{\alpha\beta\gamma\delta} u_\beta u_{\gamma;\delta} \ .
\end{equation}   

The kinematical tensor describes the difference between the Lie and
Fermi-Walker temporal derivative operators when acting on spatial
tensor fields. For example, for a spatial vector field $X$
\begin{equation}\eqalign{
     \del\fw(u) X^\alpha &
    = \del\lie(u) X^\alpha - k(u)^\alpha{}_\beta X^\beta \cr
    &= \del\lie(u) X^\alpha - \omega(u)^\alpha{}_\beta X^\beta
                         + \theta(u)^\alpha{}_\beta X^\beta   \ ,\cr
}\end{equation}   
where
\FL
\begin{equation}
    \omega(u)^\alpha{}_\beta X^\beta 
         = - \eta(u)^\alpha{}_{\beta\gamma} \omega(u)^\beta X^\gamma 
    = - [ \vec\omega(u) \times_u X ]^\alpha \ .
\end{equation}   
Spatial vector fields which undergo spatial Lie transport along $u$,
i.e., $\del\lie(u) X =0$, are called ``connecting vectors" since they have the
interpretation of being the relative position vectors of nearby observers
in the limit of vanishingly small magnitude.
This equation shows how such connecting vector fields change along $u$ with
respect to a spatial Fermi-Walker transported spatial frame along $u$,
giving the usual physical interpretation of the individual kinematical fields.
Apart from shear and expansion effects, the 
Fermi-Walker transported
spatial vectors have an angular velocity $-\vec\omega(u)$ with respect to
spatial vectors undergoing spatial Lie 
transport along $u$, or conversely the connecting
vectors rotate with angular velocity $\vec\omega(u)$ with respect
to an orthonormal spatial frame which is Fermi-Walker transported along $u$.

The kinematical quantities associated with $u$ may be used to introduce
two spacetime temporal derivatives, the Fermi-Walker 
derivative \cite{misthowhe,fer,wal} 
and the co-rotating Fermi-Walker derivative \cite{jan} 
along $u$
\FL
\begin{equation}\eqalign{
  \four\del\fw(u) X^\alpha  
     &= \four\del\sub{u} X^\alpha + [a(u)\wedge u]^{\alpha\beta} X_\beta\ ,\cr
  \four\del\cfw(u) X^\alpha  
     &= \four\del\fw(u) X^\alpha + \omega(u)^\alpha{}_\beta X^\beta\ .\cr
}\end{equation}   
These may be extended to arbitrary tensor fields in the usual way
(so that they commute with contraction and tensor products) and they both
commute with index shifting with respect to the metric and with
duality operations
on antisymmetric tensor fields since both
$\four{\rm g}$ and $\four\eta$ have zero derivative with respect to both
operators (as does $u$ itself). 
An arbitrary tensor field for which one of these operators
yields zero will be said to undergo respectively
either Fermi-Walker 
%
%
transport along $u$ or co-rotating Fermi-Walker transport along $u$.
The Fermi-Walker transport differs from parallel transport by a boost in the
plane of $u$ and $a(u)$ which maps the parallel transport of $u$ onto $u$
itself. The co-rotating Fermi-Walker transport differs by an additional
rotation in $LRS_u$ which causes it to co-rotate with the observer congruence,
i.e., to remain constant with respect to a spatial orthonormal frame undergoing
this transport, the individual frame vectors of which co-rotate with respect
to nearby observers, without undergoing the shear and expansion of the
connecting vectors.
These both differ from Lie transport along $u$ in the following  manner
\FL
\begin{equation}\eqalign{
      \Lie\sub{u} X^\alpha 
        &= \four\del\fw(u) X^\alpha
        + [ \omega(u)^\alpha{}_\beta -\theta(u)^\alpha{}_\beta 
        + u^\alpha a(u)_\beta ] X^\beta \cr
        &= \four\del\cfw(u) X^\alpha
        + [ -\theta(u)^\alpha{}_\beta 
        + u^\alpha a(u)_\beta ] X^\beta \ .\cr
}\end{equation}   

A spatial co-rotating Fermi-Walker derivative $\del\cfw(u)$
(``co-rotating Fermi-Walker temporal derivative")
 may be defined in a way analogous to the ordinary one, such that
the three temporal derivatives have
the following relation when acting on a spatial vector field $X$
\begin{equation}\eqalign{\label{eq:cfwfwlie}
  \del\cfw(u) X^\alpha 
     &= \del\fw(u) X^\alpha + \omega(u)^\alpha{}_\beta X^\beta \cr
     &= \del\lie(u) X^\alpha + \theta(u)^\alpha{}_\beta X^\beta \ ,\cr
}\end{equation}   
while $\del\cfw(u) [fu] = f a(u)$ determines its action on nonspatial fields.
It is convenient to use an index notation to handle these three operators
simultaneously
\begin{equation}\eqalign{\label{eq:temders}
&  \{ \del\tem(u) \}_{\rm tem = fw,cfw,lie} 
  =  \{ \del\fw(u), \del\cfw(u), \del\lie(u) \} \ .\cr
}\end{equation}   
The Lie temporal derivative does not commute with index shifting of spatial
fields by the metric or with the spatial duality operation using $\eta(u)$
but generates additional expansion tensor terms. Only the other two
temporal derivatives are in general compatible with imposing an 
orthonormality condition on a spatial frame which undergoes their
corresponding transport along $u$.

The restriction of the spatial Fermi-Walker derivative to purely spatial
tensor fields is the derivative first introduced by Fermi [46].
The measurement of the ordinary
or co-rotating Fermi-Walker derivative of an arbitrary tensor field
results in the corresponding spatial derivative
acting on each spatial tensor field of
the collection of fields which represent the undifferentiated tensor field.

\section{Observer-adapted frames}

Components with respect to a frame adapted to the observer orthogonal
decomposition can be quite useful in the splitting game, especially
in splitting tensor fields with many indices.
An ``observer-adapted frame" $\{e_\alpha\}$ with dual frame 
$\{\omega^\alpha\}$ will be any frame for which
$e_0$ is along $u$ and the ``spatial frame" $\{e_a\}_{a=1,2,3}$ spans the local
rest space $LRS_u$
\begin{equation}\meqalign{
  & u=  L^{-1} e_0 \equiv e_\top\ , & u^\flat (e_a)=0\ ,\cr
  & u^\flat= - L \,\omega^0 \equiv -\omega^\top\ ,\quad &  \omega^a(u)=0\ .\cr
}\end{equation}    
If it is oriented and time-oriented, then
$L>0$ and $\eta(u)_{123}>0$.
The index ``$\top$" (pronounced ``tan")
suggests ``tangential" to the congruence (or ``temporal") and
corresponds to the orthonormal temporal component obtained by scaling
the zero-indexed frame  component by the normalization factor $L$.
Similarly it is customary to use the index ``$\bot$" (``perp") in the
hypersurface point of view where $u$ is perpendicular to the integrable
distribution of local rest spaces.

The splitting of a tensor field $S$ amounts to a partitioning of the
components in an observer-adapted frame
according to whether or not individual indices are zero or not.
The purely spatial part corresponds to those components which have only
``spatial indices" 1,2,3, i.e., no ``temporal index" 0.
For a $1\choose1$-tensor $S$ one has
\begin{equation} 
S  \quad \leftrightarrow \quad
  \{ S^0{}_0,   S^a{}_0,  S^0{}_a, S^a{}_b \} \ .
\end{equation}    
Rescaling each 0 index by an appropriate factor of $L$ corresponds to the
measurement process described above, apart from the sign difference
between $u^\flat$ and $\omega^\top$.
Spatial tensors have only the spatially-indexed components nonzero,
so indexed formulas with Greek indices involving only spatial fields
reduce to Latin-indexed formulas when expressed in an observer-adapted 
frame.

The spacetime metric and its inverse in such a frame
have the form
\begin{equation}\meqalign{
\four{\rm g} 
    &= - L^2 \omega^0\otimes \omega^0 + h_{ab} \omega^a\otimes \omega^b
    &= - \omega^\top\otimes \omega^\top + h_{ab} \omega^a\otimes \omega^b
\ ,\cr
\four{\rm g}^{-1} &= -  L^{-2} e_0\otimes e_0 + h^{ab} e_a\otimes e_b
    &= - e_\top\otimes e_\top + h^{ab} e_a\otimes e_b\ ,\cr
}\end{equation}    
where $(h_{ab})$ is a positive-definite matrix
with positive determinant $h$.
The spacetime metric determinant factor has the expression
$ \four g^{1/2} =  L h^{1/2}$, while the oriented spatial volume 3-form
has components $\eta(u)_{abc} =\four\eta_{\top abc}= h^{1/2}\epsilon_{abc}$.
The spatial metric and its inverse are the covariant and contravariant
forms of the spatial projection
$  P(u) = e_a\otimes \omega^b$
\begin{equation}
  P(u)^\flat = h_{ab} \omega^a\otimes \omega^b\ , \quad
  P(u)^\sharp = h^{ab} e_a\otimes e_b\ .
\end{equation}    

One can also introduce the components of the spatial part of the spatial
connection in an observer-adapted frame by  making the usual definition
\begin{equation}
   \del(u)\sub{e_a} e_b = \Gamma(u)^c{}_{ab} e_c  \ .
\end{equation}   
Introducing several notations
$\partial_\alpha f = f_{,\alpha} = e_\alpha f$
for the frame derivatives of functions, and the anticyclic permutation
notation
\begin{equation}
    A_{\{abc\}_-} = A_{abc} - A_{bca} + A_{cab} \ ,
\end{equation}   
one finds the usual formula
\begin{equation}\label{eq:scc}
     \Gamma(u)_{abc}
     = \half [ h_{\{ab,c\}_-} + C(u)_{\{abc\}_-}  ] \ ,
\end{equation}   
where
$C(u)^a{}_{bc} = C^a{}_{bc} = \omega^a([e_b,e_c])$ 
are the spatial components of the Lie bracket 
tensor of this spatial frame with its indices
shifted from the normal positions using the spatial metric.
One then has familiar formulas like
\FL
\begin{equation}\label{eq:spacovderadaptedframe}
   [ \del(u)\sub{X} Y ]^a = X^b \del(u)_b Y^a 
    = X^b [ Y^a{}_{,b} + \Gamma(u)^a{}_{bc} Y^c ]
\end{equation}   
for two spatial vector fields $X$ and $Y$.

Of the remaining structure functions 
$C^\alpha{}_{\beta\gamma} = \omega^\alpha([e_\beta,e_\gamma])$
of the observer-adapted frame,
some are closely related to the acceleration and rotation of $u$, while the
remaining ones appear in the temporal Lie derivative of a spatial quantity,
as in
\begin{equation}\label{eq:lietemderadaptedframe}
    \del\lie(u) X^a = L^{-1} [ \partial_0 X^a + C^a{}_{0b} X^b ] \ .
\end{equation}   
This in turn leads to explicit expressions for $\del\fw(u)X$ and
$\del\cfw(u)X$ by \Eq (\ref{eq:cfwfwlie}). In particular
\begin{equation}\label{eq:temdersadaptedframe}
    \del\tem(u) e_a = C\tem(u)^b{}_a e_b \ ,
\end{equation}   
where
\begin{equation}\eqalign{\label{eq:temdersadaptedframec}
  C\lie(u)^b{}_a &=  L^{-1} C^b{}_{0a}  \ ,\cr
  C\cfw(u)^b{}_a &=  L^{-1} C^b{}_{0a} + \theta(u)^b{}_a  \ ,\cr
  C\fw(u)^b{}_a &=  L^{-1} C^b{}_{0a} + \theta(u)^b{}_a 
                                  - \omega(u)^b{}_a \ ,\cr
}\end{equation}   
indicates three useful choices for fixing the otherwise arbitrary
structure functions $C^b{}_{0a}$ which determine
how the spatial frame is transported along $u$. 
Setting the matrix $C\tem(u)^a{}_b$ to zero for each of the three choices
in turn
respectively defines the spatial frame's spatial Lie transport, 
its co-rotating Fermi-Walker transport, and its Fermi-Walker transport 
along $u$.

\section{Relative kinematics: algebra}

Suppose $U$ is another unit timelike vector field representing a different
family of test observers. One can then consider relating the ``observations"
of each to the other.
Their relative velocities are defined by
\begin{equation}\eqalign{\label{eq:boost}
   U &= \gamma(U,u) [ u + \nu(U,u) ] \ ,\qquad
   u = \gamma(u,U) [ U + \nu(u,U) ] \ ,\cr
}\end{equation}   
where the relative velocity $\nu(U,u)$ of $U$ with respect to $u$ 
is spatial with respect to $u$ and vice versa, both of which have the same
magnitude $|| \nu(U,u) ||= [ \nu(U,u)_\alpha \nu(U,u)^\alpha ]^{1/2}$, 
while the common gamma factor is related to that magnitude by
\begin{equation}
   \gamma(U,u) = \gamma(u,U) = [ 1 - ||\nu(U,u)||^2 ]^{-1/2}
         = - U_\alpha u^\alpha   
    \ .
\end{equation}   
Let $\hat\nu(U,u)$ be the unit vector giving the direction of the relative
velocity $\nu(U,u)$. 

Introduce also the energy and spatial momentum per unit mass relative to $u$
\FL
\begin{equation}
    \tilde E(U,u) = \gamma(U,u) \ , \quad \tilde p(U,u) = \gamma(U,u) \nu(U,u)
  \ .
\end{equation}   
In addition to the natural parametrization of the worldlines of $U$ by
the proper time $\tau_U$, one may introduce a new parametrization $\tau_{(U,u)}$
by
\begin{equation}\label{eq:repar}
     d \tau_{(U,u)} / d\tau_U = \gamma(U,u) \ ,
\end{equation}   
which corresponds to the sequence of
proper times of the  family of observers from the
$u$ congruence which cross paths with a given worldline of the $U$ congruence.
It is convenient to abbreviate $\gamma(U,u)$ by $\gamma$ when its meaning
is clear from the context.

\Eqs (\ref{eq:boost}) describe a unique active ``relative observer boost"
$B(U,u)$ in the ``relative observer plane"  spanned by $u$ and $U$
such that
\begin{equation}
   B(U,u) u = U\ , \quad      B(U,u) \nu(U,u) = - \nu(u,U) 
\end{equation}   
and which acts as the identity on the common subspace of the local rest spaces
$LRS_u \cap LRS_U$ orthogonal to the direction of motion. The inverse
boost $B(u,U)$ ``brings $U$ to rest" relative to $u$.
It will be convenient to use the same symbol for a linear map of the tangent
space into itself and the corresponding $1\choose1$-tensor acting by
contraction. The right contraction between two such maps will
represent their composition. When the contraction symbol is suppressed,
the linear map will be implied.

The projection $P(U)$ restricts to an invertible
map $P(U,u)=P(U)\circ P(u):LRS_u \to LRS_U$  
with inverse $P(U,u)^{-1} : LRS_U \to LRS_u$
and vice versa, and these maps also act as the
identity on the common subspace of the local rest spaces.
Similarly the boost $B(U,u)$ restricts to an invertible map
$B\lrs(U,u) \equiv P(U)\circ B(U,u)\circ P(u)$ between the
local rest spaces which also acts as the identity on their common
subspace.
The boosts and projections between the local rest spaces differ only by
a gamma factor along the direction of motion.
It is exactly the inverse projection map 
which describes Lorentz contraction of lengths along the direction of motion.
Figure 1 illustrates these maps on the relative observer plane of $u$ and
$U$.

\typeout{figure 1}

\begin{figure}[t]
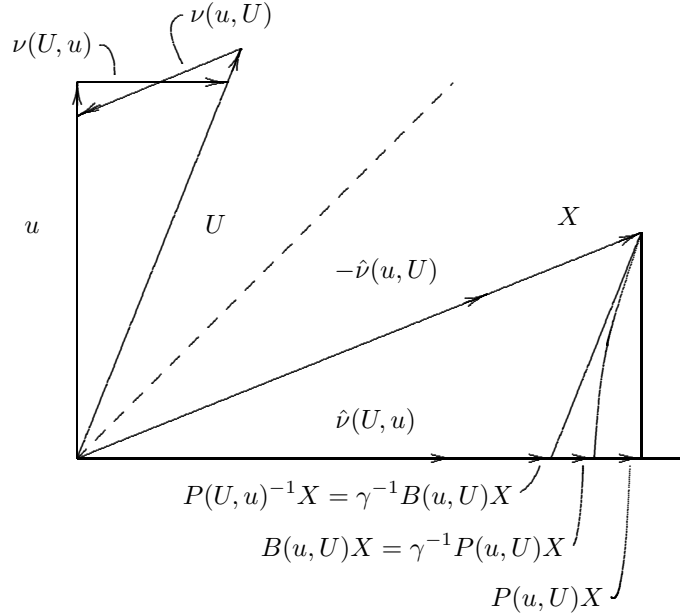

$$ \vbox{
\beginpicture
  \setcoordinatesystem units <1cm,1cm> point at 0 0  


    \putrule from 0 0   to 0 5.0 
    \putrule from 0 0   to 8.0 0  
    \putrule from 0 5.0 to 2.0 5.0 
    \putrule from 7.5 0 to 7.5 3.0 

  \setlinear
 
    \plot 0 0 2.18  5.45  /  
    \plot 0 0 7.5 3.0  /     
    \plot 0 4.55 2.18 5.45 / 
    \plot 6.3 0 7.5 3.0 /    

\setdashes

    \plot 0 0 5 5 /  


\setsolid
\setquadratic

  \plot 6.875 0  6.95 0.9  7.05 1.5  7.15 1.9  7.5 3.0 /


\setlinear
  
    \arrow <.3cm> [.1,.4]    from  0 4.7 to 0 5.0           
    \arrow <.3cm> [.1,.4]    from  2.06 5.15 to 2.18 5.45   
    \arrow <.3cm> [.1,.4]    from  1.7 5.0 to 2.0 5.0       
    \arrow <.3cm> [.1,.4]    from  0.3 4.7 to 0 4.55        
  
    \arrow <.3cm> [.1,.4]    from  4.7 0 to 5.0 0           
    \arrow <.3cm> [.1,.4]    from  5.15 2.06 to  5.45 2.18  
    \arrow <.3cm> [.1,.4]    from  7.2 2.88 to  7.5 3.0     

    \arrow <.3cm> [.1,.4]    from  6.0 0 to 6.3 0           
    \arrow <.3cm> [.1,.4]    from  6.575 0 to 6.875 0       
    \arrow <.3cm> [.1,.4]    from  7.2 0 to 7.5 0           
  \put {\mathput{u}}                          [rB]   at  -.5 3.0
  \put {\mathput{U}}                          [lB]   at  1.7 3.0
  \put {\mathput{\nu(U,u)}}                   [rB]   at  0.2 5.4
  \put {\mathput{\nu(u,U)}}                   [lB]   at  1.5 5.8

  \put {\mathput{X}}                          [rB]   at  6.7 3.1
  \put {\mathput{\hat\nu(U,u)}}               [rB]   at  4.5 0.4
  \put {\mathput{-\hat\nu(u,U)}}              [rB]   at  4.8 2.4

  \put {\mathput{P(u,U)X}}                               [rt]   at  7   -1.7
  \put {\mathput{B(u,U)X = \gamma^{-1} P(u,U)X}}         [rt]   at  6.5 -1.0
  \put {\mathput{P(U,u)^{-1}X = \gamma^{-1} B(u,U)X}}    [rt]   at  5.8 -.3

\setquadratic
  \plot 5.90 -0.45 6.05 -0.30 6.15 -0.10 /
  \plot 6.60 -1.15 6.68 -0.80 6.725 -0.10 /
  \plot 7.10 -1.85 7.25 -1.50 7.35 -0.10 /

  \plot 0.30 5.50  0.45 5.40 0.50 5.10 /
  \plot 1.40 5.90  1.25 5.70 1.40 5.25 /

\endpicture}$$
\caption{
The relationship between the various maps on the relative
observer plane of $u$ and $U$. The unit vector
$\hat\nu(U,u)$ gives the direction of the subspace belonging to
the local rest space $LRS_u$, while
$\hat\nu(u,U)$ does the same for $LRS_U$.}
\end{figure}

If $Y\in LRS_u$, then the orthogonality condition $0=u_\alpha Y^\alpha$ implies
that $Y$ has the form
\begin{equation}\label{eq:otherspdec}
   Y= [ \nu(u,U) \cdot_U P(U,u) Y ] U + P(U,u) Y \ .
\end{equation}   
If $X= P(U,u)Y \in LRS_U$ is the field seen by $U$, then $Y=P(U,u)^{-1} X$
and
\begin{equation}\eqalign{
    P(U,u)^{-1} X &= [ \nu(u,U) \cdot_U X ] U + X
           = [ P(U) + U \otimes \nu(u,U)^\flat ] \rightcontract X \ , \cr
}\end{equation}   
which gives a useful expression for the inverse projection.

This map appears in the transformation law for the electric and magnetic fields.
Suppose $\four F^\alpha{}_\beta$ is the mixed form of the
electromagnetic 2-form $\four F^\flat$. The electric and magnetic (vector)
fields
seen by $u$ result from its measurement by $u$, together with the
spatial duality operation in the latter case
\begin{equation}
   E(u) = \four F \rightcontract u \ , \quad
   B(u) = \dualp{u} P(u) \four F^\sharp \ ,
\end{equation}   
or in index notation
\FL
\begin{equation}
   E(u)^\alpha =  \four F^\alpha{}_\beta u^\beta\ , \quad
   B(u)^\alpha = \half \eta(u)^{\alpha\beta\gamma} \four F_{\beta\gamma} \ ,
\end{equation}   
with
\begin{equation}
   \four F^\flat =  u^\flat \wedge E(u)^\flat + \dualp{u} B(u)^\flat \ .
\end{equation}   

The transformation of the electric and magnetic fields is simple. For
example, using the fact that
\begin{equation}
   [ P(U) \four F ] \rightcontract \nu(u,U) = \nu(u,U) \times_U B(U) \ ,
\end{equation}   
one finds
\FL
\begin{equation}\eqalign{
  P(U,u) E(u)
       &= \gamma   P(U) \{ \four F \rightcontract [ U + \nu(u,U) ] \} \cr
      &= \gamma   [ E(U) + \nu(u,U) \times_U B(U) ] \ , \cr
}\end{equation}   
and similarly
\FL
\begin{equation}
 P(U,u) B(u) = \gamma   [ B(U) - \nu(u,U) \times_U E(U) ] \ .
\end{equation}   
Equivalently one may write
\FL
\begin{equation}\eqalign{
 E(u) &= \gamma   P(U,u)^{-1} [ E(U) + \nu(u,U) \times_U B(U) ] \ ,\cr
 B(u) &= \gamma   P(U,u)^{-1} [ B(U) - \nu(u,U) \times_U E(U) ] \ .\cr
}\end{equation}   

Any map between the local rest spaces may be ``measured" by one of 
the observers, i.e., expressed
entirely in terms of quantities which are spatial with respect to
that observer.
For example, the mixed tensor 
\begin{equation}\eqalign{
   P(U,u) &= P(U) \rightcontract P(u)
           = P(U) \rightcontract P(U,u)
           = P(U,u) \rightcontract P(u) \cr
}\end{equation}   
(which expands to $P(u) + \gamma U \otimes \nu(U,u)$),
corresponding to the linear map $P(U,u) : LRS_u \to LRS_U$, is spatial
with respect to $u$ in its covariant index and with respect to $U$
in its contravariant index, i.e., is a ``connecting tensor" in the
terminology of Schouten \cite{sch54}.
It has associated with it two tensors
\begin{equation}\eqalign{
   P(U) &= P(U,u) \rightcontract P(U,u)^{-1} \ , \cr
   P(u) &= P(U,u)^{-1} \rightcontract P(U,u) \ , \cr
}\end{equation}   
which are spatial with respect to $U$ and $u$ respectively and correspond
to identity transformations of each local rest space into itself.
In the same way any linear map $M(U,u) : LRS_u \to LRS_U$ is represented
by such a connecting tensor and has associated with it two tensors
$M_U(U,u)$ and $M_u(U,u)$ which are spatial with respect to $U$ and $u$
respectively and act as linear transformations of the respective
local rest spaces into themselves
\FL
\begin{equation}\eqalign{
    M(U,u) 
    &= M_U(U,u) \rightcontract P(U,u) \cr
            &= P(U,u) \rightcontract M_u(U,u) \ , \cr
    M_U(U,u) &= M(U,u) \rightcontract P(U,u)^{-1} \ ,\cr
    M_u(U,u) &= P(U,u)^{-1} \rightcontract M(U,u) \ .\cr
}\end{equation}   
These latter tensors enable one to express the map in terms of the spatial
projections of just one of the observers.

The individual projections parallel and perpendicular to the direction of
relative motion between the local rest spaces and within each local rest space
have the representations
\begin{equation}\imeqalign{
     P\parp(U,u) &= -\gamma  
                    \hat\nu(u,U) \otimes \hat\nu(U,u)^\flat \ ,\qquad&
    P\perpp(U,u) &= P(U,u) - P\parp(U,u) \ ,\cr
     P_U\parp(U,u) &= \hat\nu(u,U) \otimes \hat\nu(u,U)^\flat \ ,\qquad&
      P_U\perpp(U,u) &= P(U) - P_U\parp(U,u) \ ,\cr
    P_u\parp(U,u) &= \hat\nu(U,u) \otimes \hat\nu(U,u)^\flat \ ,\qquad&
      P_u\perpp(U,u) &= P(u) - P_u\parp(U,u) \ ,\cr
}\end{equation}   
where $ P(U,u) \hat\nu(u,U) = - \gamma   \hat\nu(U,u)$ explains the $\gamma$
factor in the first relation.
These in turn may be used to similarly decompose the boost $B\lrs(U,u)$ and
the inverse projection $P(u,U)^{-1}$, for which one has the obvious
relations (see Figure 1)
\FL
\begin{equation}\eqalign{\label{eq:ipbp}
     P\parp(u,U)^{-1} &= \gamma  ^{-1} B\lrs\parp (U,u)
              = \gamma  ^{-2} P\parp (U,u) \ ,\cr
     P\perpp(u,U)^{-1} &= B\lrs\perpp(U,u) = P\perpp(U,u) \cr
}\end{equation}   
which may be used to reconstruct the spatial tensors associated with the
boost and inverse projection.

For example, for the inverse boost $B\lrs(u,U)$ one has
\FL
\begin{equation}\eqalign{
    B\lrs{}_u(u,U) &= P(u) 
     - \gamma   (\gamma   + 1)^{-1} \nu(U,u) \otimes \nu(U,u)^\flat
         \ ,\cr
    B\lrs{}_U(u,U) &= P(U) 
     - \gamma   (\gamma   + 1)^{-1} \nu(u,U) \otimes \nu(u,U)^\flat
         \ ,\cr
}\end{equation}   
which follows from the expansion of
\FL
\begin{equation}\eqalign{  
    B\lrs{}_u(u,U) &= B\lrs{}\perpp{}_u(u,U) + B\lrs\parp(u,U)_u \cr
             &= P_u\perpp(u,U) + \gamma  ^{-1} P_u\parp(u,U) \ .\cr
}\end{equation}   
Thus if $S\in LRS_U$, then its inverse boost is
\FL
\begin{equation}\label{eq:spinboost}
\eqalign{ 
    B\lrs(u,U) S
    &= [ P(u) - \gamma(\gamma + 1)^{-1} \nu(U,u) \otimes \nu(U,u){}^\flat ] 
              \rightcontract P(u,U)S \ .}
\end{equation} 

The map $P(u,U) P(U,u)$ is an isomorphism of $LRS_u$ into itself which
turns up in manipulations with these maps. It and its inverse have the
following expressions
\FL
\begin{eqnarray}\label{eq:ppuu}
     P(u,U) P(U,u) 
   &=&  P_u\perpp(u) + \gamma^2 P\parp_u(u,U) \nn\\
   &=& P(u) + \gamma  ^2 \nu(U,u) \otimes \nu(U,u)^\flat \ ,\nn\\
     { [P(u,U) P(U,u)]^{-1}}
   &=& P(U,u)^{-1} P(u,U)^{-1} = P_u(U,u)^{-1} \nn\\
   &=& P_u\perpp(U,u) + \gamma  ^{-2} P\parp_u(U,u) \nn\\
   &=& P(u) - \nu(U,u) \otimes \nu(U,u)^\flat \ ,
\end{eqnarray}
giving an explicit representation of the inverse projection as well.

The transformation of the electric and magnetic fields takes a more
familiar form if one re-expresses it in terms of the parallel/perpendicular
decomposition of the boost using \Eq (\ref{eq:ipbp})
\begin{eqnarray}
    E\parp(u) &=& B\parp\lrs(u,U) E\parp(U) \ ,\nn\\
    E\perpp(u) &=& \gamma   B\perpp\lrs(u,U) [ E\perpp(U) 
                 - \nu(u,U) \times_U B\perpp(U) ] \ ,
\end{eqnarray}   
with analogous expressions for the magnetic field. When expressed in a
pair of orthonormal frames adapted to the two local rest spaces and related
by the boost, these reduce to the familiar component expressions in a
direct way.

\section{Relative kinematics: derivatives}

Suppose one uses the suggestive notation
\begin{equation}
   \four D(U) / d \tau_U = \four\del\sub{U}
\end{equation}   
for the ``total covariant derivative" along $U$. Its spatial projection 
with respect to $u$ and rescaling corresponding to the reparametrization
of \Eq (\ref{eq:repar})
is then given by the ``Fermi-Walker total spatial covariant
derivative," defined by
\FL
\begin{equation}\eqalign{
     D\fw(U,u) / d\tau_{(U,u)} &= \gamma  ^{-1} D\fw(U,u) / d\tau_U
    = \gamma  ^{-1} P(u) \four D(U) / d\tau_U \cr
   &= \del\fw(u) + \del(u)\sub{\nu(U,u)} \ .\cr
}\end{equation}   
Extend this to two other similar derivative operators
(the co-rotating Fermi-Walker and the Lie total spatial covariant derivatives)
by
\FL
\begin{equation}\eqalign{
&   D\tem(U,u) / d\tau_{(U,u)} = \del\tem(u) + \del(u)\sub{\nu(U,u)} \ ,
 \qquad {\scriptstyle {\rm tem} = {\rm fw,cfw,lie}} \ ,\cr
}\end{equation}   
which are then related to each other in the same way as the corresponding
temporal derivative operators
\begin{equation}\eqalign{\label{eq:tscds}
  D\cfw(U,u) X^\alpha /d\tau_{(U,u)}
   &= 
      D\fw(U,u) X^\alpha / d\tau_{(U,u)} + \omega(u)^\alpha{}_\beta X^\beta \cr
    &=
      D\lie(U,u) X^\alpha / d\tau_{(U,u)} + \theta(u)^\alpha{}_\beta X^\beta\cr
}\end{equation}   
when acting on a spatial vector field $X$. All of these derivative operators
reduce to the ordinary parameter derivative 
$D/d\tau_{(U,u)} \equiv d/d\tau_{(U,u)}$ when acting on
a function and extend in an obvious way to all tensor fields.
The co-rotating Fermi-Walker total spatial covariant derivative
was introduced by Massa \cite{mas,mas91}.

Explicit expressions in an observer-adapted frame
for these operators acting on spatial fields are easily obtained by
combining \Eqs (\ref{eq:spacovderadaptedframe}),
(\ref{eq:lietemderadaptedframe}) and
(\ref{eq:temdersadaptedframec}). For example, if the spatial frame
undergoes co-rotating Fermi-Walker transport along $u$, then for
a spatial vector field $X$ one finds
\begin{equation}\eqalign{\label{eq:cfwtscd}
 &  D\cfw(U,u) X^a / d \tau_{(U,u)}  
   = d X^a / d \tau_{(U,u)}  + \Gamma(u)^a{}_{bc} X^c \nu(U,u)^b \ .\cr
}\end{equation}   

Introduce the ordinary and co-rotating Fermi-Walker and the Lie
``relative accelerations" of $U$ with respect to $u$ by
\begin{equation}\eqalign{\label{eq:relacc}
   &   a\tem(U,u) = D\tem(U,u) \nu(U,u) / d\tau_{(U,u)} \ ,
   \qquad {\scriptstyle {\rm tem} = {\rm fw}, {\rm cfw}, {\rm lie}} \ .\cr
}\end{equation}   
These are related to each other in the same way as the corresponding
derivative operators in \Eq (\ref{eq:cfwfwlie}).

The total spatial covariant derivative 
operators restrict in a natural way to a single timelike 
worldline with 4-velocity $U$, where the $D/d\tau$ notation is most 
appropriate; $\four D(U) / d\tau_U$ is often called the
absolute or intrinsic derivative along the worldline of $U$
(associated with an induced connection along such a worldline \cite{sacwu}).
One can then study a single worldline of a test particle
with respect to the given family of test observers.
One can also introduce corresponding spatial transport operations along
$U$ using these three operations by requiring that a field have zero
derivative along the worldline with respect to the corresponding
total spatial covariant derivative.
Call these spatial ordinary and co-rotating Fermi-Walker transport and spatial
Lie transport along $U$ with respect to $u$, where the term
``spatial" is understood to refer to $u$.

\section{Spatial gravitational fields}

The worldline of a test particle of nonzero mass $m$, 
timelike 4-velocity $U$ and
4-acceleration $a(U)$
under the influence of a force $f(U)$ satisfies
the ``acceleration equals force" equation
\begin{equation}
   a(U) = \tilde f(U) \equiv f(U) / m
\end{equation}   
which equates the 4-acceleration to the force per unit mass.
The spatial projection of this equation with respect to $u$ generalizes
the more familiar spatial force equation of special relativity by the
occurrence of kinematical terms which may be interpreted as ``spatial
gravitational forces," as well as by the presence of the spatial covariant
derivative. Before projecting this equation, one must define the spatial
acceleration and force analogous to their special relativistic definitions.

According to \Eq (\ref{eq:otherspdec}) with $u$ and $U$ interchanged,
any vector $X$ which is orthogonal to $U$ has the following decomposition
with respect to $u$
\begin{equation}
    X =   [ \nu(U,u) \cdot_u P(u,U)X ] u + P(u,U) X \ .
\end{equation}   
In  particular the acceleration 
$a(U) = \four D(U) U / d\tau_U$
of the unit vector $U$ and
therefore the applied 4-force per unit  mass 
$\tilde f(U)$ are orthogonal to $U$ so
\FL
\begin{equation}\eqalign{
  a(U)   &= \gamma(U,u) [ \nu(U,u) \cdot_u  A(U,u) \, u +  A(U,u) ] \ ,\cr
  \tilde f(U) &= \gamma(U,u) [ \nu(U,u) \cdot_u \tilde F(U,u) \, u 
                       + \tilde F(U,u) ] \ ,\cr
}\end{equation}   
where
\begin{equation}\eqalign{
    A(U,u) &= \gamma(U,u)^{-1} P(u,U) a(U) \ ,\cr
   \tilde F(U,u) &= \gamma(U,u)^{-1} P(u,U) \tilde f(U) \ .\cr
}\end{equation}   

Evaluating the rescaled spatial projection $ A(U,u)$ of the acceleration
by making explicit the projection of the total covariant derivative of $U$
\FL
\begin{equation}\eqalign{
  A(U,u) &= \gamma(U,u)^{-1} P(u) \four D(U) U / d\tau_U \cr
         &= D\fw(U,u) [ \gamma(U,u) u + \tilde p(U,u) ] / d\tau_{(U,u)} \cr
       &= D\fw(U,u) \tilde p(U,u) / d\tau_{(U,u)} - \tilde F{}\fw\g(U,u) \ ,\cr
}\end{equation}  
where
\begin{equation}\label{eq:fwsgf}
   \tilde F{}\fw\g(U,u) = - \gamma(U,u) D\fw(U,u) u / d\tau_{(U,u)} \ ,
\end{equation}  
leads to the introduction of three analogously defined quantities
\FL
\begin{equation}\label{eq:temsgf}
\eqalign{
  A(U,u) 
       &= D\tem(U,u) \tilde p(U,u) / d\tau_{(U,u)} 
                       - \tilde F{}\tem\g(U,u) \ ,
   \quad {\scriptstyle {\rm tem} = {\rm fw,cfw,lie}} \ ,}
\end{equation} 
which are related to each other
in the same way that the corresponding total spatial
covariant derivatives (and relative accelerations when $X=u$)
are related to each other in \Eq (\ref{eq:tscds}).
One may also express these relations in terms of the relative accelerations
by substituting $\tilde p(U,u) = \gamma(U,u) \nu(U,u)$, leading to
\FL
\begin{equation}\eqalign{
   A(U,u) 
      &= \gamma   P(u,U)\rightcontract P(U,u) \rightcontract a\tem(U,u)
           - \tilde F\g\tem(U,u)  \ , 
 \quad {\scriptstyle {\rm tem} = {\rm fw,cfw}} \ ,\cr
}\end{equation}  
which can be inverted to yield
\FL
\begin{equation}\label{eq:relacctwo}
\eqalign{ 
   a\tem(U,u) 
    &= \gamma(U,u)^{-1} P_u(U,u)^{-1} \rightcontract
       [ \tilde F\g\tem(U,u) +  A(U,u) ]  \ , 
   \quad {\scriptstyle {\rm tem} = {\rm fw,cfw}} \ ,}
\end{equation} 
where the composed projection map and the relative
spatial projection tensor are given by \Eq (\ref{eq:ppuu}).
An additional expansion term along $\nu(U,u)$
arises in the Lie case from the derivative of $\gamma(U,u)$.

Given these definitions coming from analyzing the acceleration alone,
the rescaled spatial projection $A(U,u) = \tilde F(U,u)$
of the force equation $a(U) = \tilde f(U)$ can
then be expressed in the form 
\FL
\begin{equation}\label{eq:spaforceeq}
\eqalign{
    D\tem(U,u) \tilde p(U,u) / d\tau_{(U,u)} 
     &= \tilde F\g\tem(U,u) + \tilde F(U,u) \ , 
  \quad {\scriptstyle {\rm tem} = {\rm fw,cfw,lie}} \ ,}
\end{equation} 
leading to the
identification of the terms
$\tilde F\g\fw(U,u)$,  $\tilde F\g\cfw(U,u)$,  and $\tilde F\g\lie(U,u)$ 
respectively as the
ordinary and co-rotating Fermi-Walker and the Lie spatial gravitational forces
per unit mass.
Since index shifting does not commute with the Lie total spatial covariant
derivative,
it is convenient to define also the covariant Lie spatial gravitational
force $\tilde F\g\lief(U,u)$ per unit mass by
\FL
\begin{equation}\label{eq:liefsgf}
 [ D\lie(U,u) \tilde p(U,u)^\flat  / d\tau_{(U,u)} ]^\sharp 
      = \tilde F\g\lief(U,u) + \tilde F(U,u) \ .
\end{equation}   
Similarly the rescaled temporal projection of the force
equation yields the power equation
\begin{equation}\eqalign{\label{eq:power}
  D (U,u) \tilde E(U,u) /d\tau_{(U,u)}     
    &= [ \tilde F\g\tem(U,u) + \tilde F(U,u) ] \cdot_u \nu(U,u) \cr &\quad\quad
        + \epsilon\tem  \gamma(U,u) \theta(u)^\flat(\nu(U,u),\nu(U,u)) \ ,\cr
&\quad
 \epsilon\tem = (0,0,1,-1) \ , 
 \quad {\scriptstyle {\rm tem} = {\rm fw,cfw,lie,lie\flat}} \ .\cr
}\end{equation}    

Explicit expressions for the various spatial gravitational forces
follow from their definitions (\ref{eq:fwsgf}), (\ref{eq:temsgf}), and
(\ref{eq:liefsgf}).
The covariant Lie total spatial covariant derivative differs from
the Lie total spatial covariant derivative by an expansion term arising
from the commutation of the index shifting and the derivative. All of these
forces have the same general form
\FL
\begin{equation}
     \tilde F\g\tem(U,u) 
      = \gamma(U,u) [ \vec g(u) + H\tem(U,u) \rightcontract \nu(U,u) ] \ ,
\end{equation}   
where
\begin{equation}
      \vec g(u) = - a(u)
\end{equation}   
defines the gravitoelectric vector field   $g(u)^\alpha$  and
\begin{equation}\eqalign{
     H\fw(u) &= \omega(u) - \theta(u) = k(u) \ ,\cr
     H\cfw(u) &= 2\omega(u) - \theta(u)        \ ,\cr
     H\lie(u) &= 2\omega(u) - 2\theta(u) = 2k(u) \ ,\cr
     H\lief(u) &= 2\omega(u) \ ,\cr
}\end{equation}   
define the various mixed
gravitomagnetic tensor fields  $ H\tem(u)^\alpha{}_\beta$ 
that may be introduced.
If one defines a single gravitomagnetic vector field  $ H(u)^\alpha$  by
\begin{equation}
     \vec H(u) = 2 \vec\omega(u) \ ,
\end{equation}   
then the antisymmetric part of the gravitomagnetic force contributes a term
\FL
\begin{eqnarray}
 && \ALT  H\tem(u) \rightcontract \nu(U,u) = \varepsilon\tem \,
    \nu(U,u) \times_u \vec H(u)  \ ,\nn\\
    && \quad \varepsilon\tem = (\half,1,1,1)
    \ ,\quad {\scriptstyle {\rm tem} = {\rm fw,cfw,lie,lie\flat}} \ ,
\end{eqnarray}
with the term in the Fermi-Walker spatial gravitational force
differing by a factor of one half from those of the remaining
gravitational forces.

For comparison the spatial force associated with the
electromagnetic Lorentz force on a test particle
of charge $q$ and  mass $m$ due to electric and magnetic fields
$E(u)$ and $B(u)$ measured by $u$ 
and the corresponding spatial gravitational force are
is
\FL 
\begin{eqnarray}
       F\em(u) &=& q [ E(u) + \nu(U,u) \times_u B(u) ] \ ,\nn\\
     F\g\tem(U,u) &=& m \gamma(U,u) [ \vec g(u) + 
        \varepsilon\tem \, \nu(U,u) \times_u \vec H(u) \nn\\
       &&\quad \quad + \SYM H\tem(u) \rightcontract \nu(U,u) ] \ ,\nn\\
    && \quad \quad {\scriptstyle {\rm tem} = {\rm fw,cfw,lie,lie\flat}} \ .
\end{eqnarray}
Apart from the additional gamma factor in the spatial gravitational force
and the symmetric tensor contribution,
the close similarity of the two  expressions
makes the analogy with the Lorentz force and the origin of the
gravitoelectromagnetic jargon clear.

The gravitomagnetic symmetric tensor field,
which has no analog in electromagnetism, arises from the temporal
derivative of the spatial metric, which is the new ingredient. The spatial
derivatives of the spatial metric
enter the total spatial covariant derivative of the spatial
momentum as a ``space curvature" force term
\FL
\begin{equation}\eqalign{\label{eq:lietscd}
   D\tem(U,u) \tilde p(U,u)^a /  d \tau_{(U,u)} 
      &=  d \tilde p(U,u)^a / d \tau_{(U,u)} 
             +  C\tem(u)^a{}_b \tilde p(U,u)^b \cr
     &\quad\quad + \Gamma(u)^a{}_{bc} \tilde p(U,u)^c \nu(U,u)^b \cr
}\end{equation}   
which is quadratic in the spatial velocity. One can thus think of the spatial
metric as a potential for  these two different spatial forces.
The matrix $C\tem(u)^a{}_b$ given by \Eq (\ref{eq:temdersadaptedframec})
depends on how the spatial frame is transported along the congruence
and may be conveniently be chosen to vanish for one of the three temporal
derivatives.

To handle the case of a lightlike test particle with zero rest mass,
one must work with the null 4-momentum $P$ instead of a 4-velocity.
The substitutions
\begin{equation}\eqalign{
 &      (U,\gamma(U,u),\tilde p(U,u),\nu(U,u))
         \to (P,E(P,u),p(P,u),\nu(P,u)) \cr
}\end{equation}   
lead to analogous spatial force and power equations for this case.

\section{Maxwell-like equations}

The analogy between the gravitoelectromagnetic vector fields and the
electromagnetic ones
shows that the exterior derivative of the observer
velocity 1-form corresponds to the electromagnetic 2-form, which is
itself locally the exterior derivative of a 4-potential 1-form
\begin{equation}\eqalign{\label{eq:obsgpot}
     d \four A &= u^\flat \wedge E(u)^\flat + \dualp{u} B(u)^\flat 
                                            = \four F^\flat \ ,\cr
     d u^\flat &= u^\flat \wedge \vec g(u)^\flat 
         + \dualp{u} \vec H(u)^\flat \ .\cr
}\end{equation}   
The observer 4-velocity thus acts as the 4-potential for the
gravitoelectromagnetic vector fields. 

The splitting of the
identity $d^2 \four A^\flat =0$ leads to half of Maxwell's equations
\begin{equation}\eqalign{\label{eq:meq}
    & \div_u B(u) + \vec H(u) \cdot_u E(u) = 0 \ , \cr
    & \curl_u E(u) - \vec g(u) \times_u E(u) 
             + [ \Lie(u)\sub{u} + \Theta(u) ] B(u) =0 \ .\cr
}\end{equation}   
Replacing $\four A$ by $\four u$ reduces these to the corresponding
gravitoelectromagnetic equations
\begin{equation}\eqalign{\label{eq:gemeq}
    & [ \div_u + \vec g(u) \cdot_u ] \vec H(u) = 0 \ , \cr
    & \curl_u \vec g(u) + [ \Lie(u)\sub{u} + \Theta(u) ] \vec H(u) =0 \ .\cr
}\end{equation}   

Splitting the remaining half of Maxwell's equations
\begin{equation}
    \dual d \dual \, \four F = 4\pi \four J
\end{equation}   
leads to
\begin{equation}\eqalign{\label{eq:meqsources}
    & \div_u E(u) - \vec H(u) \cdot_u B(u) = 4\pi \rho(u) \ , \cr
    & \curl_u B(u) - \vec g(u) \times_u B(u) 
             -  [ \Lie(u)\sub{u} + \Theta(u) ] E(u) = 4\pi J(u) \ ,\cr
}\end{equation}   
where $  \four J = \rho(u) u + J(u) $
is the splitting of the 4-current.

The remaining Maxwell-like equations for the gravitoelectromagnetic
vector fields arise from the Einstein equations.
In order to state them one must first introduce appropriate spatial
curvature tensors associated with
the spatial part of the spatial connection of the observer congruence $u$.
There are in fact four different spatial
curvature tensors one may introduce \cite{jancar91}.
Three of them have the invariant definition
\FL
\begin{eqnarray}
 &&\big\{ [ \del(u)\sub{X}, \del(u)\sub{Y}] -\del(u)\sub{[X,Y]}\big\} Z \nn\\
   &&\quad= R\tem(u)(X,Y) Z
        + 2\omega(u)^\flat(X,Y) \del\tem(u) Z\ , \nn\\
   && \quad\quad \quad  {\scriptstyle {\rm tem} \,=\, {\rm fw, cfw, lie } } \ ,
\end{eqnarray} 
where $X$, $Y$, and $Z$ are spatial vector fields.  These three tensors,
the Fermi-Walker spatial curvature tensor \cite{mas},
the co-rotating Fermi-Walker spatial curvature tensor \cite{jancar91}
and the Lie spatial curvature tensor \cite{zel,cat58}
differ by the same kinematical terms as the temporal derivative operators
themselves but reversed in sign and in a tensor product with twice the
rotation tensor.
In an observer-adapted frame their components are
\FL
\begin{equation}\eqalign{
    R\tem(u)^a{}_{bcd}
&= 2 \partial_{[c} \Gamma(u)^a{}_{d]b} - C^e{}_{cd} \Gamma(u)^a{}_{eb}\cr
&\qquad + 2 \Gamma(u)^a{}_{[c|e|} \Gamma(u)^e{}_{d]b}
          - 2 C\tem(u)^a{}_b \omega(u)_{cd} \ .\cr
}\end{equation}   

The fourth ``symmetry-obeying" spatial curvature 
tensor \cite{ferr65,jancar91}
$R\sym(u)$
is related to them by the following component formulas in an observer-adapted
frame
\FL
\begin{equation}\eqalign{
 R\sym (u)^{ab}{}_{cd} &= R\lie(u)^{ab}{}_{cd}
   - 2 \theta(u)^{ab} \omega(u)_{cd} 
   -4 \theta(u)^{[a}{}_{[c} \omega(u)^{b]}{}_{d]} \cr
&= R\lie(u)^{[ab]}{}_{cd}
   -4 \theta(u)^{[a}{}_{[c} \omega(u)^{b]}{}_{d]} \cr
&= R\cfw(u)^{[ab]}{}_{cd}
   -4 \theta(u)^{[a}{}_{[c} \omega(u)^{b]}{}_{d]} \cr
&= R\fw(u)^{[ab]}{}_{cd}
   - 2 \omega(u)^{ab} \omega(u)_{cd}  
   -4 \theta(u)^{[a}{}_{[c} \omega(u)^{b]}{}_{d]} \cr
}\end{equation}   
and obeys all the usual symmetry identities of a 3-metric curvature
tensor, as discussed by  
Ferrarese \cite{ferr65},
so one may define its symmetric Ricci tensor $R\sym(u)^a{}_b$
and symmetric Einstein tensor $G\sym(u)^a{}_b$ by the usual
formulas.
For a hypersurface splitting in which $u$ has vanishing vorticity,
all of these spatial curvature tensors
coincide with the curvature tensor of the induced Riemannian metric on the
spacelike hypersurfaces orthogonal to $u$.

The spacetime Einstein tensor, the scalar curvature and the spacetime
Ricci tensor have the
following components in an observer-adapted frame in each
point of view
\FL
\begin{equation}\eqalign{\label{eq:einstein} 
2 \four G^\top{}_\top &= 
     \Tr \theta(u)^2 - \Theta(u)^2 
     - \fraction32  H(u)^c  H(u)_c - R\sym(u)^c{}_c \ ,\cr
2 \four G^\top{}_a &= 2 \four R^\top{}_a
    = -  2\del(u)_b[ \theta(u)^{b}{}_a - \delta^b{}_a \Theta(u)]
  - \{[\vec \del(u) - 2 \vec g(u)] \times_u \vec H(u)\}_a\ ,\cr
\four G^a{}_b &=   \{ \Lie(u)\sub{u} +\Theta(u) \}
       [ \theta(u)^a{}_b -\delta^a{}_b \Theta(u)] 
   + \half \delta^a{}_b [\Tr \theta(u)^2 - \Theta(u)^2] \cr
  &\quad 
     + [\del(u)_{(b}  - g(u)_{(b}]g(u)^{a)} 
      - \delta^a{}_b [\del(u)_c - g(u)_c] g(u)^c 
      \cr & \quad   
     - \half H(u)^a H(u)_b  +\fraction14 \delta^a{}_b H(u)^c H(u)_c
     + G\sym(u)^a{}_b \ ,\cr
\four R &= 2 \{\Lie(u)\sub{u} + \Theta(u)\} \Theta(u)
     + \Tr \theta(u)^2 -\Theta(u)^2 
     + 2[\del(u)_a - g(u)_a] g(u)^a 
      \cr & \quad  
     + \half H(u)^c H(u)_c
     + R\sym(u)^c{}_c \ ,\cr
 \four R^\top{}_\top &=
    [\Lie(u)\sub{u} +\Theta(u)] \Theta(u) +\Tr \theta(u)^2 -\Theta(u)^2
      \cr & \quad  
   + [ \del(u)_c - g(u)_c] g(u)^c
   - \half H(u)^c H(u)_c \ ,\cr
\four R^a{}_b &=   -\{ \Lie(u)\sub{u} + \Theta(u) \} \theta(u)^a{}_b
    + [\del(u)_{(b} - g(u)_{(b}] g(u)^{a)}
      \cr & \quad  
    - \half [H(u)^a H(u)_b - \delta^a{}_b H(u)^c H(u)_c]
     + R\sym(u)^a{}_b\ .\cr
}\end{equation}    
The spatial scalar and spatial vector equations which result from the
measurement of the Ricci form of the Einstein equations
provide the much more complicated analogs of the source driven pair
of Maxwell equations
\begin{equation}\eqalign{
     \four R^\top{}_\top 
          &=  8\pi [ \four T^\top{}_\top - \half\four T^\alpha{}_\alpha ]\ ,\cr
   2 \four R^\top{}_a &= 16\pi \four T^\top{}_a \ ,\cr
}\end{equation}   
The complication arises from the new ingredient described by the spatial metric
which has no analog in linear electrodynamics. This effect (``space curvature")
also appears  in the
acceleration equals force equation in the total spatial covariant derivative
operator.

\section{Transformation of spatial gravitational fields}

The spatial gravitational force fields are simply related to the kinematical
quantities associated with the observer congruence. If one has two
distinct observer congruences with unit tangents $u$ and $U$, one
can describe the transformation law between the spatial gravitational
fields observed by each. One need only express the quantities and operators
in the expression for the spatial gravitational fields of one in terms of those
of the other to obtain such laws, as in the above derivation of the 
transformation law for the electric and magnetic fields.

Abbreviating $\gamma(U,u)$ to $\gamma$,
the acceleration and kinematical field transform as follows
\begin{equation}\eqalign{
    a(U) &= \gamma^2 P(u,U)^{-1} [ a(u) - k(u) \rightcontract \nu(U,u) ] 
              \cr&\quad 
               + \gamma^2 P(U,u) a\fw(U,u)  \ ,\cr
   k(U) &= \gamma^2 P(U,u) [ k(u) - a(u) \otimes \nu(U,u)^\flat ] 
               \cr&\quad 
             - \gamma \del(U) \nu(U,u) \ .\cr
}\end{equation}   
The rotation tensor and vector then transform as
\FL
\begin{equation}\eqalign{
    \omega(U)^\flat &= \gamma^2 P(U,u) 
           [ \omega(u)^\flat - \half a(u)\wedge \nu(U,u) ]
               \cr&\quad 
           + \half \gamma d(U) \nu(U,u)^\flat \ ,\cr
   \vec\omega(U) &= \gamma^2 P(u,U)^{-1}
           [ \vec\omega(u) + \half\nu(U,u)\times_u a(u) ]
               \cr&\quad 
           + \half \gamma \curl_U \nu(U,u) \ ,\cr
}\end{equation}   

Converting to the gravitoelectromagnetic symbols leads to
\FL
\begin{equation}\eqalign{
   \vec H(U) &= \gamma^2 P(u,U)^{-1}
           [ \vec H(u) - \nu(U,u)\times_u \vec g(u) ]
               \cr&\quad 
           + \gamma \curl_U \nu(U,u) \ ,\cr
    \vec g(U) &= \gamma^2 P(u,U)^{-1} 
         [ \vec g(u) + \half \nu(U,u) \times_u \vec H(u) 
               \cr&\quad 
           - \theta(u) \rightcontract \nu(U,u) ]
               - \gamma^2 P(U,u) a\fw(U,u)  \cr
       &=  \gamma^2 P(u,U)^{-1} 
           [ \vec g(u) + \nu(U,u) \times_u \vec H(u) 
               \cr&\quad 
           - \theta(u) \rightcontract \nu(U,u) ]
               - \gamma^2 P(U,u) a\cfw(U,u)  \ ,\cr
}\end{equation}   
where the expressions in square brackets in the gravitoelectric field
transformation laws are just $\gamma^{-1} \tilde F{}\g\fw(U,u)$ and
$\gamma^{-1} \tilde F{}\g\cfw(U,u)$ respectively,
analogous to the Lorentz force and its magnetic analog which appear in the
transformation law for the electric and magnetic fields.
The terms explicitly involving the gravitoelectromagnetic
vector fields
in the transformation law for the 
gravitomagnetic vector field and in the second form of the one for the
gravitoelectric vector field
are exactly analogous to the
corresponding transformation laws for the magnetic and electric fields,
apart from the extra gamma factor
also present in the force law itself.

Apart from the expansion term,
the remaining part of the transformation law which breaks this
correspondence, namely the
relative acceleration and the relative velocity curl, 
can be further expanded. The relative acceleration, for example,
can be re-expressed using \Eq (\ref{eq:relacctwo}) 
with $A(U,u)$ replaced by $\tilde F(U,u)$.
For the relative curl, one can apply the following useful formula
\begin{equation}\eqalign{
   \curl_U X 
      &= \gamma P(u,U)^{-1} \{ \curl_u X 
        + \nu(U,u) \times_u [\Lie(u)\sub{u} X^\flat ]^\sharp \} \ ,\cr
}\end{equation}   
valid when $X$ is spatial with respect to $u$.

For an observer-adapted co-rotating Fermi-Walker orthonormal frame,
the gravitoelectric and gravitomagnetic fields are related
to the 2-form which results from evaluating the
tensor-valued connection 1-form on $u$ in a way similar to the way
the electric and magnetic fields are related to the electromagnetic
2-form \cite{jancar91}.
The homogeneous part of the transformation law for  the
connection then leads to the terms in the transformation law for the
gravitoelectric and gravitomagnetic vector fields which are
analogous to those for the electric and magnetic fields.

\section{Gyroscope precession}

The spin vector of a test gyro carried by an observer of the observer 
congruence undergoes Fermi-Walker transport along $u$ and therefore
rotates relative to a co-rotating Fermi-Walker transported spatial frame with
an angular velocity of ``gravitomagnetic precession" given by
\begin{equation}\eqalign{
   D\cfw(u,u) S / d\tau_u &= \zeta\gmag(u) \times_u S \ ,\qquad
      \zeta\gmag(u) = -\vec\omega(u) = -\half \vec H(u) \ .\cr
}\end{equation}   

Along an arbitrary timelike worldline with 4-velocity $U$, the spin vector
has the decomposition
\begin{equation}
    S = [ \nu(U,u) \cdot_u \vec S ] u + \vec S\ , \quad
       \vec S \equiv P(u,U) S 
\end{equation}   
and its length $|| S || = [S_\alpha S^\alpha]^{1/2}$ remains constant under
Fermi-Walker transport. The spin vector $\vec S$ observed by $u$
both rotates and changes in magnitude.

By straightforward projection of the Fermi-Walker transport equation, one
finds for the observed spin vector
\FL
\begin{eqnarray}
D\cfw  (U,u) \vec S /d\tau_{(U,u)}
 &=&   -\vec\omega(u) \times_u \vec S +
  \gamma^{-1} [ \nu(U,u) \cdot_u \vec S ] \tilde F{}\g\fw(U,u)  \nn\\
  &&
  -  \gamma [A(U,u)^\flat \rightcontract P_u(U,u)^{-1} \rightcontract \vec S ]
        \,\nu(U,u)   \ ,
\end{eqnarray}    
where the relative
spatial projection tensor is given by \Eq (\ref{eq:ppuu}).

Introducing its length $||\vec S|| = [\vec S \cdot_u \vec S]^{1/2}$
and its direction $\hat S = ||\vec S||^{-1} \vec S$, one finds
\FL
\begin{equation}\eqalign{ 
  D\cfw (U,u)  \ln || \vec S || /d\tau_{(U,u)}
    &= \gamma^{-1} [ \nu(U,u) \cdot_u \hat S ] \,
                [ \tilde F{}\g\fw(U,u)  \cdot_u \hat S] \cr
 &\qquad
  -  \gamma [A(U,u)^\flat \rightcontract P_u(U,u)^{-1} \rightcontract \hat S ]
       \, [\nu(U,u)  \cdot_u \hat S ] \ ,\cr
}\end{equation}   
and
\begin{equation}
  D\cfw(U,u) \hat{S} /d\tau_{(U,u)} 
             = \Omega\cfw(\hat S,U,u) \times_u \hat{S} \ ,
\end{equation}   
where
\FL
\begin{equation}\eqalign{ 
   \Omega\cfw(\hat S,U,u)
      &=   - \vec\omega(u) 
     -  \gamma^{-1} [ \nu(U,u) \cdot_u \hat{S} ]  \, 
          \tilde F{}\g\fw(U,u) \times_u \hat{S} \cr
      & \qquad
  +  \gamma [A(U,u)^\flat \rightcontract P_u(U,u)^{-1} \rightcontract \hat{S} ]
      \, \nu(U,u) \times_u \hat{S} \ .\cr
}\end{equation}   
These formulas describe the precession of the spin vector as seen by
the family of different observers of the observer congruence along the
gyro's worldline.

To describe the precession as seen by the observer carrying the gyro,
one must first decide with respect to what the precession will be measured.
Suppose $\{e_a\}$ is an orthonormal spatial frame which is tied to the 
congruence, i.e., undergoes co-rotating Fermi-Walker transport along $u$.
The observer carrying the gyro will see these axes at each event along his
worldline to be in relative motion. The orientation of these moving axes
with respect to the local rest space of this observer can only be defined
to be the  orientation of the axes which are aligned with the moving axes
but momentarily at rest. (In fact, the projections $P(U,u)e_a$ into $LRS_U$,
namely the moving frame vectors as seen by $U$,
will no longer be orthonormal.) Thus the orientation of the spin vector $S$
with respect to the axes $B\lrs(U,u)e_a$ 
``momentarily at rest" is well-defined
and represents the orientation of $S$ with respect to the moving axes $e_a$.

However, the orientation of $S$ with respect to $B\lrs(U,u)e_a$ is the same as
the orientation with respect to $e_a$ of the spin vector 
${\cal S} \equiv B\lrs(u,U)S$ momentarily at rest with respect to the congruence
observer, since the boost is an isometry.
Thus the angular velocity of the spin vector with respect to the sequence
of congruence spatial frames as observed
in its own local rest space equals the angular velocity of the boosted
spin vector  relative to the sequence of congruence frames 
as observed by the sequence of congruence observers,
apart from a proper time renormalization.

The boosted spin vector ${\cal S}$ is given by \Eq (\ref{eq:spinboost})
rewritten in terms of the momentum per unit mass
\begin{equation}
   {\cal S}   = \vec S
         - \gamma^{-1}(\gamma + 1)^{-1} 
                 [\tilde p(U,u) \cdot_u \vec S] \tilde p(U,u)  
\end{equation}    
and its evolution along the worldline is a simple consequence of the
above result for $\vec S$ 
coupled with the spatial force equation for $\tilde p(U,u)$
and the power equation for $\tilde E(U,u) = \gamma $.
The result 
\begin{equation}
    D\tem(U,u) {\cal S} / d\tau_{(U,u)} 
          = \zeta\tem(U,u) \times_u {\cal S} \ ,
    \qquad  {\scriptstyle {\rm tem} \,=\, {\rm fw, cfw} }
\end{equation}
defines respectively the Fermi-Walker and
co-rotating Fermi-Walker ``relative angular
velocities" $\zeta\fw(U,u)$ and 
\begin{equation}
  \zeta\cfw(U,u) = -\vec\omega(u) + \zeta\fw(U,u)
\end{equation}   
of $U$ with respect to $u$.
The latter one
may be expressed in the form
\FL
\begin{equation}\eqalign{
   \zeta\cfw(U,u) &= -\half \vec H(u) 
        - \gamma   (\gamma   +1)^{-1}      \nu(U,u) \times_u \tilde F(U,u) \cr
    &\qquad  + (\gamma   + 1 )^{-1} \, \nu(U,u) \times_u \tilde F\g\fw(U,u) \cr
   &= \zeta\gmag(u) + \zeta\thom(U,u) + \zeta\geo(U,u) \cr
}\end{equation}   
in terms of the spatial projection $\tilde F(U,u)$ of the applied force or
equivalently
\FL
\begin{eqnarray}
   \zeta\cfw(U,u) &=& - \half\vec H(u)
   - \gamma  ^2 (\gamma   + 1 )^{-1}  \nu(U,u) \times_u  a\cfw(U,u) \nn\\
   && + \nu(U,u) \times_u [ \tilde F\g\cfw(U,u)  \nn\\
   && \qquad - \gamma (\gamma   + 1 )^{-1} \nu(U,u) \times_u \vec\omega(u) ]
\end{eqnarray}
in terms of the relative acceleration, the latter formula due to
Massa and Zordan \cite{maszor}.
This angular velocity, in contrast with the result for $\Omega(\hat S,U,u)$,
depends only on the relative boost between the local rest spaces.
One
may show that in terms of the corresponding connecting
tensor field, the Fermi-Walker relative angular velocity
is given by the following Lie algebra type derivative expression
\begin{equation}
  \{ P(u)  [ \four D(U) \, B\lrs(u,U)/ d\tau_{(U,u)} 
        \rightcontract B\lrs(U,u) ] \}^\sharp
                        = - \dualp{u} \zeta\fw(U,u) \ .
\end{equation}

The co-rotating Fermi-Walker relative angular velocity describes how
the boosted spin vector rotates with respect to an orthonormal spatial
frame defined along the gyro worldline by spatial co-rotating Fermi-Walker
transport (spatial in each case with respect to the observer congruence).
This transport transports an orthonormal spatial frame parallel to itself
except for the additional boost which keeps it spatial and the minimal
rotation needed to keep it from rotating with respect to the observer
congruence.  However, if the worldline returns to a given observer
of the observer congruence, the spatial curvature of the spatial part of
the spatial connection $\del(u)$ will result in a net rotation compared
to that observer relative to his neighbors. 

If one is really interested
in measuring the rotation relative to a preferred orthonormal spatial
frame tied to the congruence by spatial co-rotating Fermi-Walker transport,
then one must eliminate the additional rotation due to the spatial curvature.
For such an orthonormal spatial frame, 
the expression (\ref{eq:cfwtscd}) for the co-rotating
Fermi-Walker total spatial covariant derivative along $U$ 
decomposes into the ordinary derivative minus
a term due to the relative rotation of a moving spatial co-rotating
Fermi-Walker transported frame and the one at rest in the congruence
\begin{equation}\eqalign{
   D\cfw(U,u)  {\cal S}^a / d\tau_{(U,u)}
   &    = d  {\cal S}^a / d\tau_{(U,u)} 
           + \Gamma(u)^a{}_{bc} \nu(U,u)^b {\cal S}^c \cr
  &     = d  {\cal S}^a / d\tau_{(U,u)} - 
       \eta(u)^a{}_{bc} \zeta\sc(U,u){}^b {\cal S}^c \ ,\cr
}\end{equation}   
The ``space curvature" angular velocity $\zeta\sc(U,u)$
is just the spatial dual of the
value of the tensor-valued
antisymmetric spatial connection 1-form evaluated along the
relative velocity.

One then has the angular velocity of the boosted spin vector relative to
the preferred frame as the sum of the space curvature term associated
with the preferred frame
and the relative angular velocity of the spin vector
with respect to the observer congruence. In terms of components in such
a frame one has
\FL
\begin{equation}
   d {\cal S}^a / d\tau_{(U,u)}  =
       \epsilon_{abc} [\zeta\cfw(U,u) + \zeta\sc(U,u)]{}^b {\cal S}^c \ .
\end{equation}   

The original angular velocity of the spin vector relative to the
preferred frame as seen by the observer carrying the gyro then has the
expression
\begin{equation}\label{eq:frameangvel}
     \zeta(U,u,e) = \gamma [ \zeta\cfw(U,u) + \zeta\sc(U,u) ] \ ,
\end{equation}   
taking into account the relative proper time factor. 

The co-rotating Fermi-Walker relative angular velocity consists of
three terms. The first term, the gravitomagnetic precession,
is also referred to as the
frame-dragging or Lense-Thirring \cite{lenthi} precession,
and is independent of the relative velocity of the gyro.
The last two terms together, containing explicit
factors of the relative velocity, define the Fermi-Walker relative
angular velocity $\zeta\fw(U,u)$.
The first of these two terms,
due to nongravitational forces (or possible Riemann tensor forces),
is the Thomas precession \cite{tho},
while the remaining term is called the geodetic or de Sitter or Fokker
precession \cite{des,fok,pir}.

In flat spacetime with $u$ a unit timelike Killing vector field,
corresponding to time translations,
the spatial gravitational force is zero and
all three relative accelerations coincide with the ``usual" 3-acceleration
of special relativity.
The second term in either formula for the co-rotating Fermi-Walker
angular velocity 
\FL
\begin{equation}\eqalign{
   \zeta\thom(U,u) & =
    - \gamma  ^2 (\gamma   + 1 )^{-1}  \nu(U,u) \times_u  a\cfw(U,u) \cr
& = - ||\nu(U,u)||^{-2} ( \gamma   - 1) \nu(U,u) \times_u a\cfw(U,u) \cr
}\end{equation}   
is the Thomas precession due to the acceleration of the gyro.
For circular motion with angular velocity $\vec\Omega$,
the precession angular velocity is $[\gamma  -1]\vec\Omega$, as described in
exercise 6.9 of Misner, Thorne and Wheeler \cite{misthowhe}.
The first term in their equation 6.28 is exactly the boosted
spin vector ${\cal S}$. In the limit $||\nu(U,u)|| \ll 1$ and $\gamma\to1$
of nonrelativistic motion, the Thomas precession reduces to
\begin{equation}\eqalign{
    \zeta\thom(U,u) &\rightarrow -\half \nu(U,u) \times_u a\cfw(U,u) \cr
   &\rightarrow - \half \nu(U,u) \times_u \tilde F(U,u) \ .\cr
}\end{equation}   
 
For a geodesic in an arbitrary spacetime, the Thomas precession
vanishes leaving the last term which describes the geodetic precession.
In the limit of nonrelativistic motion, it reduces to
\begin{equation}
    \zeta\so(U,u) = \half \nu(U,u) \times_u \vec g(u) \ .
\end{equation}   
Thorne \cite{tho88} describes this nonrelativistic term as an
``induced gravitomagnetic precession" or ``spin-orbit" precession
since it corresponds to the gravitomagnetic precession due to
an additional ``induced" gravitomagnetic field
$\vec H(u)\ind = - \nu(U,u) \times_u \vec g(u)$ induced by the motion
of the gyro in the gravitoelectric field in analogy with the induced
magnetic field due to motion in an electric field.

\section{Spatial gravitational potentials}

The action of the gravitational field on test bodies is described in the
context of a partial splitting of spacetime by the spatial gravitational force
fields and the spatial part of the spatial connection, all of which
together represent the spacetime connection.
In a certain sense the spatial metric $P(u)^\flat$ is a potential for the
spatial part of the spatial connection $\del(u) \circ P(u)$ and 
for the expansion tensor, while 
$u^\flat$ itself acts as a potential for the vector spatial gravitational
force fields through \Eq (\ref{eq:obsgpot}).
However, the latter relationship does not involve the
spatial or temporal derivatives of spatial quantities, like the scalar
and vector potentials that result from the splitting of the electromagnetic
4-potential. 
One needs a full splitting in order
to introduce a 4-potential for the gravitoelectromagnetic vector fields
in a way analogous to the electromagnetic case.

Suppose one has a parametrized nonlinear reference frame as described in the
introduction. This may be specified locally by a pair $(e_0,\omega^0)$
consisting of the differential $\omega^0 = dt$ of some time function for the
slicing and a vector field $e_0$ tangent to the threading with
$\omega^0(e_0) =1$, so that in a comoving coordinates with respect to
$e_0$ (i.e., local coordinates $\{x^\alpha\} = \{t,x^a\}$ ``adapted"
to the parametrized nonlinear reference frame),
$e_0$ has the representation
$\partial/\partial t$. In the slicing and threading
points of view, $\omega^0$ and $e_0$ respectively are timelike,
determining the temporal features of the nonlinear reference frame through
the specification of the observer congruence, while the remaining element
of the pair determines the choice of spatial gauge (the threading in the
slicing point of view and the slicing in the threading point of view).

In the slicing point of view, the slicing 1-form is timelike and can
be normalized, while in the threading point of view
the threading vector field is timelike and
can be normalized
\FL
\begin{equation}\imeqalign{
 \omega^{\bot} &= N \omega^0\ ,\quad &
  N^{-2} &=-\four{\rm g}^{-1}(\omega^0,\omega^0)\ ,\cr
 e_\top &= M^{-1} e_0 \equiv m\ ,\quad &
  M^{2} &=-\four{\rm g}(e_0,e_0)\ ,\cr
}\end{equation}    
and then index-shifted and sign-reversed to define
the dual object
\begin{equation}
  e_{\bot}  = -\omega^{\top\,\sharp} \equiv n\ , \quad
  \omega^\top = - e_\top{}^\flat \ .
\end{equation}   
The future-pointing
unit normal $n\equiv e_{\bot}$
to the slicing
and the unit tangent vector field $m\equiv e_\top$ to the threading
respectively serve as the 4-velocity of the corresponding family
of test observers for the two points of view, which will be referred to
commonly as $o$. The normalization factors $L(n)=N$ and $L(m)=M$,
the lapse functions in each point of view,
relate the observer proper time along the observer worldlines
to the parametrization associated with
the parametrized nonlinear reference frame
\begin{equation}
       d \tau_n / dt = N\ ,\quad
       d \tau_m / dt = M\ .
\end{equation}   

Splitting the spatial gauge field then yields the shift vector field $\vec N$
and 1-form $\overeq M$ respectively as its spatial projection
\begin{equation} 
\meqalign{
 e_0&= T(n)e_0 + P(n)e_0     
       &= N e_\bot + \vec N \ ,\cr
 \omega^0 &= T(m)\omega^0 + P(m)\omega^0 
  &= M^{-1} \omega^\top + \overeq M\ .\cr
}\end{equation}    
In slicing point of view the shift is most naturally considered as
a vector field,
and it determines the tilting of the threading curves away
from the normal direction $n$.
In the threading point of view the shift is most naturally considered as
a 1-form, 
and it determines the tilting of the threading local rest spaces
$LRS_m$ away from the directions tangent to the slicing.
Let $\overeq N= \vec N {}^\flat$ and $\vec M=\overeq M{}^\sharp$
denote the slicing shift 1-form and the threading 
shift vector field respectively.
The lapse and shift terminology in the slicing point of view is due
to Wheeler \cite{whe}.

In the slicing point of view, $N^{-1} \vec N$ is the relative velocity
of the threading with respect to the observer congruence, 
while in the threading point of view $M \vec M$ is the 
negative of the
relative velocity
of the direction normal to the slicing with respect to the observer
congruence. When the slicing is spacelike and the threading timelike,
both points of view hold and these equal the quantities $\nu(m,n)$ and
$-\nu(n,m)$ respectively associated with the relative observer boost $B(m,n)$,
which has gamma factor $\gamma(m,n) = N/M = d\tau_{(n,m)}/d\tau_m$. Both
relative velocities (and shifts)
vanish in the case of an orthogonal nonlinear reference frame
where the slicing and threading points of view coincide.
One can also introduce the terminology ``quasi-orthogonal"
for a nonlinear reference frame for which
both points of view and the condition $||\nu(m,n)||\ll 1$ are valid.

The nonlinear reference frame determines a ``reference decomposition"
of each tangent space into the threading subspace along $e_0$ 
(projection ${\cal T}$) and into the
slicing subspace (projection ${\cal P}$)
which is the kernel of $\omega^0$. The representation
of the orthogonal observer decomposition of the tangent space in terms of
this (in general) nonorthogonal decomposition provides potentials for the
spatial gravitational force fields.
In the case in which both the slicing and threading points of view hold,
the restriction of the reference spatial projection ${\cal P}$
to $LRS_m$ gives the inverse map $P(m,n)^{-1}$ associated with the
relative observer boost, while its restriction to the dual $LRS_n^\ast$
of the slicing local rest space gives the inverse map $P(n,m)^{-1}$.

Complete the pair $(e_0,\omega^0)$ to an adapted frame $\{ e_\alpha \}$ with
dual frame $\{ \omega^\alpha \}$, where the spatial frame $\{ e_a \}$
spans the slicing subspace of the tangent space at each point. Choose the
spatial frame to be comoving, i.e., Lie dragged along $e_0$. Then it is
a computational frame as introduced by York \cite{yor},
characterized by only having its spatial structure functions nonzero
\begin{equation}\eqalign{
  & C^\alpha{}_{\beta\gamma} 
      = \omega^\alpha([e_\beta,e_\gamma])
     = \delta^\alpha{}_a \delta^b{}_\beta \delta^c{}_\gamma C^a{}_{bc} \ .\cr
}\end{equation}   
Finally let $ g_{ab} = P(n)_{ab}$ and $\gamma_{ab} = P(m)_{ab}$
respectively
denote the spatial metric components in this frame, following the conventions
of Misner, Thorne and Wheeler \cite{misthowhe} and 
Landau and Lifshitz \cite{lanlif75} respectively.

Given these definitions,
the spacetime metric 
$\four{\rm g} = \four g_{\alpha\beta}\omega^\alpha\otimes\omega^\beta$
and inverse metric 
$\four{\rm g}{}^{-1} = \four g^{\alpha\beta}e_\alpha\otimes e_\beta$
in the computational
frame in the slicing point of view are
\FL
\begin{eqnarray}
 \four{\rm g}  
   &=& -N^2\omega^0\otimes\omega^0
   +g_{ab}(\omega^a +N^a\omega^0)\otimes(\omega^b +N^b\omega^0) \nn\\
&\equiv& -N^2 \omega^0\otimes\omega^0 +g_{ab}\theta^a\otimes\theta^b\ ,\nn\\
 \four{\rm g}{}^{-1}  
  &=&  -N^{-2}(e_0-N^ae_a)\otimes (e_0-N^be_b)
   +g^{ab}e_a \otimes e_b\nn\\
&\equiv& -N^{-2} \epsilon_0\otimes\epsilon_0 +g^{ab} e_a\otimes e_b
\ ,
\end{eqnarray} 
i.e., in components
\FL
\begin{equation}\imeqalign{ 
 \four g_{00}&=-(N^2-N_cN^c)\ ,\qquad &  
    \four g^{00}&=-N^{-2}\ ,\cr
 \four g_{0a} &=N_a\ , \qquad &
    \four g^{0a} &=N^{-2}N^a\ ,\cr
  \four g_{ab} &=g_{ab}\ , \qquad &
  \four g^{ab} &=g^{ab}-N^{-2}N^aN^b\ ,\cr
}\end{equation}    
where $(g^{ab})$ is the matrix inverse of the positive-definite
matrix $(g_{ab})$.
The single independent component of the volume 4-form, i.e., the 
(absolute value of the) square root of the metric determinant, is
$\four g^{1/2} = N g^{1/2}$.

In the threading point of view they are instead given by
\FL
\begin{eqnarray}
 \four{\rm g}&=&    -M^2(\omega^0 -M_a\omega^a)\otimes(\omega^0 -M_b\omega^b)
   +\gamma_{ab}\omega^a \otimes \omega^b \nn\\
&\equiv& -M^2 \theta^0\otimes \theta^0 + \gamma_{ab}\omega^a\otimes\omega^b
\ ,\nn\\
 \four{\rm g}{}^{-1}&=& -M^{-2}e_0 \otimes e_0
+\gamma^{ab}(e_a+M_ae_0) \otimes (e_b+M_be_0)\nn\\
&\equiv& -M^{-2}e_0\otimes e_0 +\gamma^{ab}\epsilon_a\otimes\epsilon_b
\ ,
\end{eqnarray} 
i.e., in components
\begin{equation}\imeqalign{ 
 \four g_{00} &= -M^2\ ,\qquad &
                \four g^{00} &=-(M^{-2}-M_cM^c)\ ,\cr
 \four g_{0a} &=M^2 M_a\ ,\qquad &
                \four g^{0a} &= M^a\ ,\cr
 \four g_{ab} &=\gamma_{ab}- M^2 M_aM_b\ , \qquad &
  \four g^{ab} &=\gamma^{ab}\ .\cr
}\end{equation}
Here
the spatial metric matrix $(\gamma_{ab})$ is positive-definite, with
inverse $(\gamma^{ab})$. 
Letting $\gamma=\det(\gamma_{ab})>0$, one has
$\four g^{1/2} =M\gamma^{1/2}$.

In the threading point of view the projected computational frame
$\{ e_0,\epsilon_a \}$ with dual frame $\{ \theta^0, \omega^a \}$,
where $\theta^0 = T(m)\omega^0$ and $\epsilon_a = P(m)e_a$, is an
observer-adapted frame which is also ``spatially-comoving," namely
it undergoes spatial Lie transport along the observer congruence which
coincides with the threading.
In the slicing point of view the projected computational frame
$\{ \epsilon_0,e_a \}$ with dual frame $\{ \omega^0, \theta^a \}$,
where $\epsilon_0 = T(n)e_0$ and $\theta^a = P(n)\omega^a$, is an
observer-adapted frame, but it is not spatially-comoving along the 
observer congruence. Instead it undergoes spatial (with respect to $n$)
Lie transport along the threading.

In both points of view, spatial fields have only spa\-tial\-ly-indexed 
com\-pu\-ta\-tion\-al frame
components nonzero and their indices may be shifted using the
spatial metric component matrices. The reference decomposition of a
tensor field corresponds to the partition of computational components
according the reference
``temporal" index 0 and the 
reference ``spatial " indices $1,2,3$. 
Covariant (contravariant) spatial indices correspond to the slicing
(threading) spatial projection, while
contravariant (covariant) temporal indices correspond to the slicing
(threading) temporal projection.

For the slicing and threading
parametrizations of the spacetime metric, it is
precisely the two explicit terms in the final representation of $\four{\rm g}$
and $\four{\rm g}^{-1}$ above
which correspond to the covariant and contravariant form of the
orthogonal projections along the local time and space directions.
The spatial metric in each case is just the covariant form of the
projection. In the slicing point of view, its restriction to a slice
yields the induced metric on the slice submanifold making it into
a Riemannian manifold,
while in the threading
point of view it yields the projected metric on the slice submanifold,
making the slice into a different Riemannian manifold representing the
projected geometry rather than the induced geometry, the latter of which is not
necessarily Riemannian in the threading point of view without an
additional causality assumption.
In each point of view, however,
this spatial metric describes the relative distances of the
worldlines of nearby observers at a given coordinate time $t$.

The parametrized nonlinear reference frame enables one to
represent the spatial tensor algebra over spacetime in a natural way
in terms of the
tensor algebra of time-dependent tensor fields over the ``computational
3-space," namely the quotient space of the spacetime by the threading
congruence. On this 3-manifold, the time-dependent spatial metric is
a Riemannian metric and its connection may be related to the spatial
part of the spatial connection  by a difference tensor. In the slicing
point of view this difference tensor is zero since the spatial metric
and the spatial connection
corresponds to the induced metric on each slice with its own associated
connection.  In the threading point of view it involves the expansion
tensor of the observer congruence.

Let $\four\Gamma^\alpha{}_{\beta\gamma} 
= \omega^\alpha(\four\del\sub{e_\beta} e_\gamma)$ be the computational
components of the spacetime connection, and let
$\Gamma(o)^a{}_{bc}$ be the components of the spatial part of the spatial
connection in the spatial projected computational frame,
as defined by \Eq (\ref{eq:scc}).
Then the components of the spatial part of the spatial connection
are
\FL
\begin{eqnarray}
   \Gamma(n)_{abc} &=& \four \Gamma_{abc} 
     = \half [ g_{\{ab,c\}_-} + C(n)_{\{abc\}_-}  ] \ ,\nn\\
   \Gamma(m)_{abc} 
       &=& \gamma_{ad} \,\gamma_{be} \gamma_{cf} \four \Gamma^{def} \nn\\
      &=& \half [ \gamma_{\{ab,c\}_-} + \partial_0 \gamma_{\{ab}M_{c\}_-}
    + C(m)_{\{abc\}_-}  ] \nn\\
      &=& \Gamma(\gamma)_{abc} + M \theta(m)_{ \{ab } M_{c\}_-} \ ,
\end{eqnarray}
where $\Gamma(\gamma)^a{}_{bc}$ is the connection of the projected metric
on the slice, corresponding to the connection of the time-dependent metric
on the computational 3-space.
The expansion tensor terms arise in the threading point of view 
since the observer-adapted spatial frame derivatives $\epsilon_a f 
=\partial_a f + M_a \partial_0 f$ which occur in expressing spatial
components of $d(m) f$ also involve the reference time derivative.

One may introduce all of the spatial gravitational force fields associated with
the observer congruence. The threading point of view is just
a representation of the congruence point of view associated with $m$
by expressing it in terms of the nonlinear reference frame, so it provides
potentials for those fields. 
However,
the slicing point of view is distinct from
the hypersurface point of view associated with $n$ since it describes 
evolution in terms of the threading rather than the normal congruence
and so employs a temporal derivative along $e_0$ rather than along the
observer congruence. One must introduce a corresponding Lie total spatial
covariant derivative and redefine the spatial gravitational force by
the difference term. This reintroduces a gravitomagnetic force in the
slicing point of view which is zero in the hypersurface point of view
due to the vanishing vorticity of the normal congruence.

The Lie temporal derivatives associated with the observer congruence are
\begin{equation}\eqalign{
   \del\lie(m) &= M^{-1} \Lie(m)\sub{e_0} \ ,\cr
   \del\lie(n) &= \del\lie(n,e_0) - N^{-1} \Lie(n)\sub{\vec N} \ ,\cr
}\end{equation}   
where $ \del\lie(n,e_0) \equiv   N^{-1} \Lie(n)\sub{e_0} $
defines the slicing point of view Lie temporal derivative, which
is used in that point of view in order to measure evolution along the
threading congruence.  The temporal
derivatives $M\del\lie(m)$ and $N\del\lie(n,e_0)$ of a spatial tensor field
on spacetime
just reduce to the ordinary time derivative of the corresponding
time-dependent tensor field on the computational 3-space.
The kinematical tensor $k(n)$ is just the extrinsic curvature in the
slicing point of view, which has the familiar form
\begin{equation}\eqalign{
    k(n)_{ab} &= - \theta(n)_{ab}
    = - \half N^{-1} [ \Lie(n)\sub{e_0} g_{ab} - 2 \del(n)_{(a} N_{b)} ] \cr
}\end{equation}   
when written in terms of the slicing Lie temporal derivative.

In the slicing point of view the various total spatial covariant
derivatives all correspond to derivatives along the vector field
\begin{equation}\eqalign{
    \gamma(U,n)^{-1} U
         &= n + \nu(U,n) \cr
         &= N^{-1} e_0 + [ \nu(U,n) - N^{-1} \vec N ] \ ,\cr
}\end{equation}   
the latter form of which is its reference decomposition.
The slicing point of view, measuring evolution with respect to the
nonlinear reference frame, implements this reference decomposition
with the new temporal derivative along the temporal component 
\FL
\begin{equation}\eqalign{
 D\lie(U,n,e_0) X / d\tau_{(U,n)}
    &= \del\lie(n,e_0) X 
               + \del(n)\sub{[\nu(U,n) - N^{-1}\vec N]} X \cr
      &= D\lie(U,n) X / d\tau_{(n,U)} 
               - \Delta H\lie(n,e_0) \rightcontract X \ ,\cr
}\end{equation}   
leading to the difference term
\FL
\begin{equation}\eqalign{
  \Delta H\lie(n,e_0) \rightcontract X 
      &= N^{-1} [ \del(n)\sub{\vec N} - \Lie(n)\sub{\vec N} ] X \ ,\cr
  [\Delta H\lie(n,e_0)]^\alpha{}_\beta 
      &= N^{-1} [\del(n) \vec N ]^\alpha{}_\beta
      = N^{-1} \del(n)_\beta  N^\alpha \ ,\cr
}\end{equation}   
which must  be added to the hypersurface Lie spatial gravitational force
to obtain the slicing version.

The various Lie gravitomagnetic tensors are
\FL
\begin{equation}\imeqalign{
     H\lief(m)_{\alpha\beta} &= 2 M  \del(m)_{[\alpha} M_{\beta]} \ , \quad&
     H\lie(m)_{\alpha\beta} &=     H\lief(m)_{\alpha\beta}
          -  M^{-1} \Lie(m)\sub{e_0} \gamma_{\alpha\beta} \ , \cr
     H\lief(n)_{\alpha\beta} &= 0  \ ,\quad&
     H\lie(n)_{\alpha\beta} 
         &= -  N^{-1} \Lie(n)\sub{e_0-\vec N} g_{\alpha\beta}
          = -2 \theta(n)_{\alpha\beta} \ ,\cr
     H\lief(n,e_0)_{\alpha\beta} &= N^{-1} \del(n)_\alpha N_\beta\ , \quad&
     H\lie(n,e_0)_{\alpha\beta} &=  H\lief(n,e_0)_{\alpha\beta}
                 -  N^{-1} \Lie(n)\sub{e_0} g_{\alpha\beta} \ ,\cr
}\end{equation}   
where $g_{\alpha\beta}=P(n)_{\alpha\beta}$ and
$\gamma_{\alpha\beta}=P(m)_{\alpha\beta}$ and
only Latin indices are necessary
when expressed in the projected computational frame.
The gravitomagnetic vectors are
\begin{equation}\eqalign{
   \vec H\lie(m) &= M \curl_m \vec M \ , \cr
   \vec H\lie(n) &= 0\ , \cr
   \vec H\lie(n,e_0) &= N^{-1} \curl_n \vec N \ ,\cr
}\end{equation}   
and the gravitoelectric vectors are
\begin{equation}\eqalign{
   \vec g(m) &= -\grad_m ( \ln M) - [\Lie(m)\sub{e_0} \overeq M ]^\sharp\ ,\cr
   \vec g(n) &= -\grad_n ( \ln N) \ .\cr
}\end{equation}   
This leads to the interpretation of
\begin{equation}
      \Phi(o) = \ln L(o)
\end{equation}   
as the scalar gravitational potential and the shift as the vector
gravitational potential in each point of view, together determining the vector
gravitational force fields.
With these definitions
the slicing Lie total spatial gravitational forces are then
\begin{equation}\eqalign{
  \tilde F{}\g\tem(U,n,e_0) 
  &= \gamma(U,n) [ \vec g(n) 
       + \half \nu(U,n) \times_n \vec H\lie(n,e_0)  \cr &\qquad
+  \SYM H\tem(n,e_0)\rightcontract \nu(U,n) ] \ .\cr
    &\quad \quad {\scriptstyle {\rm tem} = {\rm lie,lie\flat}} \ ,\cr
}\end{equation}   

The threading point of view is merely a representation of the congruence
point of view associated with $m$, so 
\begin{equation}
   m^\flat = - \omega^\top = - M [ \omega^0 - \overeq M ]
\end{equation}   
acts as the 4-potential of the gravitoelectromagnetic vector fields
as in \Eq (\ref{eq:obsgpot}).
A similar statement describes the hypersurface point of view, but
the slicing point of view does not admit a 4-potential in this sense.

The gravitoelectric and gravitomagnetic force fields have been discussed in
the black hole case in the slicing point of view
using the slicing total spatial covariant derivative operator 
by Thorne et al \cite{thoprimac}.
Both Zel'manov \cite{zel} and Cattaneo 
\cite{cat58}--\cite{cat61} 
have discussed them
from the threading point of view, while Landau and 
Lifshitz \cite{lanlif41,lanlif75}
discuss only the stationary case in the threading point of view.
M\o ller \cite{mol} 
discusses them both from his 
parametrization-dependent description of the threading point of view
as well as for the threading point of view. All of these threading point
of view discussions introduce the covariant Lie spatial gravitational forces.
Massa \cite{mas} has re-expressed the Cattaneo approach in a
somewhat more modern framework, using the 
co-rotating Fermi-Walker total spatial covariant derivative.

The gravitoelectromagnetic terminology is apparently
due to Thorne and seems to have
evolved from Forward's linearized discussion \cite{for} of M\o ller's 
threading point of view work 
(Forward uses the term protational for gravitomagnetic)
which Forward
used to draw an analogy between the electromagnetic field and the
linearized gravitational field in general relativity.
This
generalized to the parametrized-post-Newtonian (PPN) discussion of 
Braginsky, Caves and Thorne \cite{bracavtho} where the terminology
``electric-type" and ``magnetic-type" gravitational fields appeared,
the latter of which became the ``gravitomagnetic field" in a 
post-Newtonian
general relativistic 
discussion of Braginsky, Polnarev and Thorne \cite{brapoltho}.
The ``gravitoelectric field" and the gravitomagnetic tensor force
finally appeared in the context of black holes in the slicing point of view
in the book by Thorne et al \cite{thoprimac}.

Having introduced potentials for the gravitoelectric and gravitomagnetic
vector fields by a choice of parametrized nonlinear reference frame,
one can discuss the effect of spatial gauge transformations.
In the threading point of view one can change the slicing, keeping the
threading fixed, which will obviously not effect any quantitites defined
only in terms of the threading decomposition. In particular the
gravitoelectric and gravitomagnetic fields will be invariant,
leading to a gauge freedom analogous to that of the scalar and vector
potentials for the electric and magnetic fields.

The slicing point of view is a hybrid which depends both on the
slicing and the threading, so changing the threading will leave invariant
only those fields associated with the corresponding hypersurface point of
view. The gravitomagnetic vector field will change, although the
gravitoelectric field is trivially invariant since the lapse function
is invariant.

Perhaps the easiest way to discuss these transformations is via an
adapted coordinate system $\{t,x^a\}$. If one changes the slicing
by the following change of adapted coordinates
\begin{equation}
    t = t( t', x') \ , \quad x^a = x'{}^a \ ,
\end{equation}   
then one easily finds from the new form of the metric that
\begin{equation}\eqalign{
     M'   &= [\partial_0' t] M \ ,\cr
     M_a' &= [\partial_0' t]^{-1} [ M_a - \partial_a' t ]  .\cr
}\end{equation}   
The change of threading for fixed slicing is less interesting. The
lapse is invariant and the shift is augmented by an additional vector
field which is the difference between the old and new shift vector fields,
apart from the change of spatial coordinates which is induced.

One can also consider the temporal gauge freedom associated with
changing the observer congruence itself in each point of view.
In the slicing (threading) point of view, this corresponds
to a change of slicing (threading). The change of slicing leads to
an similar tranformation of the lapse and shift as well as the spatial metric
in the slicing point of view.

\section{Second-order acceleration equation}

In the context of a nonlinear reference frame which enables
one to represent the spacetime geometry in terms of time-dependent
fields on a computational 3-space, one can re-express the 
first order spatial acceleration equation for the spatial momentum
as a second-order
equation describing the evolution of the spatial coordinates along the
worldline under consideration. It is exactly this equation that
describes the Coriolis and centrifugal forces in rotating coordinates
in flat spacetime and which is necessary to interpret the spatial
force equation in the context of an adapted coordinate system in
actual problems of interest.

Suppose $\{x^\alpha\} =\{t,x^a\}$ 
are local coordinates adapted to the parametrized
nonlinear reference frame, i.e., comoving with respect to $e_0$, so that
the coordinate frame is a computational frame, with $C^a{}_{bc}=0$.
Let ${\cal U}^\alpha = d x^\alpha / dt \equiv \dot x{}^\alpha$ be the coordinate
components of the coordinate velocity of a worldline with 4-velocity $U$
\FL
\begin{equation}
     U^\alpha = d x^\alpha/ d\tau_U = \gamma(U,o) dx^\alpha/ d\tau_{(o,U)}
                      = \Gamma(U,o) \dot x{}^\alpha \ ,
\end{equation}   
where
\begin{equation}
   \Gamma(U,o) \equiv  |{\cal U}_\alpha {\cal U}^\alpha |^{-1/2} 
        = dt/d\tau_U \ ,
\end{equation}   
which has the respective values
\FL
\begin{eqnarray}
    \Gamma(U,m)
     &=& M^{-1}[ (1- M_a \dot x{}^a )^2 
                     - M^{-2}\gamma_{ab} \dot x{}^a \dot x{}^b ]^{-1/2} \nn\\
                  &=& M^{-1} \gamma(U,m) (1-M_a \dot x{}^a)^{-1} \ ,\nn\\
    \Gamma(U,n)
 &=& N^{-1} [ 1-  N^{-2}g_{ab} (\dot x{}^a + N^a)( \dot x{}^b + N^b)]^{-1/2} 
       \nn\\
                  &=& N^{-1} \gamma(U,n) \ ,
\end{eqnarray}
is M\o ller's coordinate gamma factor expressed in the two points of view.
Its sign-reversed
reciprocal is the coordinate time Lagrangian for the timelike geodesics
\begin{equation}
   I = - {\scriptstyle\int} \, d\tau_U 
     = - {\scriptstyle\int} \Gamma(U,o)^{-1} \, dt \ .
\end{equation}   
The momenta canonically conjugate to $x^a$ are
\begin{equation}
    \pi_a = \left\{
  \eqalign{ & \Gamma(U,n) g_{ab} ( \dot x{}^b + N^b ) \cr
          & \Gamma(U,m) \gamma_{ab} \dot x{}^b + \tilde {\cal E}(U,m)M_a  \cr}
            \right.
\end{equation}   
leading to the Hamiltonian $H$ which equals the
``coordinate energy" (per unit mass)
$\tilde {\cal E}(U,o) = - U_0$ which has the respective expressions
\FL
\begin{equation}\eqalign{ 
 \tilde{\cal E}(U,m) &=
   \gamma(U,m) M = M^2 \Gamma(U,m) (1-M_a \dot x{}^a) \ ,\cr
 \tilde{\cal E}(U,n) &=
         \gamma(U,n) N ( 1 - \nu(U,n) \cdot_n N^{-1}\vec N ) \cr
           &= N^2 \Gamma(U,n) (1- N^{-2} N_a [\dot x{}^a + N^a] ) \ .\cr
}\end{equation}   

The rate of change of the coordinate time with respect to the observer proper
time parametrization of the worldline with 4-velocity $U$ is
\FL
\begin{equation}\eqalign{
      dt/d\tau_{(o,U)} &= { dt/d\tau_U \over d\tau_{(o,U)}/d\tau_U} \cr
              &= \Gamma(U,o)/\gamma(U,o) 
          = \left\{ \eqalign{ &M^{-1} (1-M_a \dot x{}^a)^{-1} \cr
                              &N^{-1} \cr} \right. \ . \cr
}\end{equation}   
Note that in the threading point of view, the rates of change of
coordinate time with respect to the observer proper time differ on the
observer worldlines and the general worldline, while they agree in the
slicing point of view.

The rates of change $\dot x{}^a$ of the spatial coordinates
with respect to the coordinate time $t$ define the 
reference spatial components of the coordinate spatial velocity
\begin{equation}\label{eq:coordrelmom}
   {\cal U}(U,o)^a =   
  \left\{\meqalign{ & \Gamma(U,m)^{-1} \tilde p(U,m)^a \cr
                    & \Gamma(U,n)^{-1} \tilde p(U,n)^a - N^a \cr
                      }\right.
\end{equation}   
or in terms of the velocity
\begin{equation}\label{eq:coordrelvel}
   {\cal U}(U,o)^a =   
  \left\{\eqalign{ 
                         & M ( 1- M_b \dot x{}^b )  \nu(U,m)^a\cr
                         & N \nu(U,n)^a - N^a   \ , \cr
                   }\right.
\end{equation}     
which in turn
define a spatial vector in each point of view
\FL
\begin{equation}
  {\cal U}(U,n) = {\cal U}(U,n)^a e_a \ ,\quad
   {\cal U}(U,m) = {\cal U}(U,m)^a \epsilon_a \ .
\end{equation}   

The projected computational frame components of the 
appropriate Lie total spatial covariant derivative of the spatial
vector ${\cal U}(U,o)$
yield the corresponding
second derivatives of the spatial coordinates
\begin{equation}\eqalign{
  \left(\matrix{ D\lie(U,m)^2 x^a /  dt^2 \cr 
                 D\lie(U,n,e_0)^2 x^a /  dt^2 \cr
      }\right) &= 
  \left(\matrix{ D\lie(U,m) {\cal U}(U,m)^a /  dt \cr 
                 D\lie(U,n,e_0) {\cal U}(U,n)^a /  dt \cr
      }\right) \cr
  &= \frac{d^2 x^a}{dt^2} 
         + \Gamma(o)^a{}_{bc} \frac{dx^b}{dt} \frac{dx^c}{dt} \ .
}\end{equation}   
These may be evaluated by insertion of the relation (\ref{eq:coordrelmom})
between spatial momentum and coordinate velocity 
into the spatial force \Eq (\ref{eq:spaforceeq}).

In the threading point of view one has M\o ller's result \cite{mol}
\begin{equation}\eqalign{ 
      \Gamma & (U,m)^{-1}  D\lie(U,m)  [ \Gamma(U,m) \dot x{}^a ] / dt \cr
         &= M^2 (1-M_b \dot x{}^b)^2
           \gamma(U,m)^{-1} D\lie(U,m) \tilde p(U,m)^a / d\tau_{(U,m)} \cr
    &= (1-M_b\dot x{}^b)^2
     [ -\grad_m \, \half M^2 - M^2 \{ \Lie(m)\sub{e_0} \overeq M \}^\sharp 
       + M^2 \gamma(U,m)^{-1}  \tilde F(U,m) ]^a \cr
   &\quad  + (1-M_b\dot x{}^b) [ M^2 {\cal U}(U,m) \times_m \curl_m \vec M 
      + M \SYM H(m) \rightcontract {\cal U} ]^a \ .\cr
}\end{equation}    
In the slicing point of view, something more interesting happens because
of the additional shift term in the coordinate relative velocity
\begin{equation}\eqalign{
  \Gamma (U,n)^{-1}  & D\lie(U,n,e_0) [ \Gamma(U,n) ( \dot x{}^a + N^a) ] / dt
      \cr & \quad  
   = N^2 \gamma(U,n)^{-1} 
         D\lie(U,n,e_0) \tilde p(U,n)^a / d\tau_{(n,U)} \cr
}\end{equation}   
leading to
\begin{equation}\eqalign{ 
     \Gamma(U,n)^{-1}  & D\lie(U,n,e_0) [ \Gamma(U,n) \dot x{}^a ] / dt \cr
   &\quad= [ -\grad_n \, \half N^2 -  \{ \Lie(n)\sub{e_0} \overeq N \}^\sharp 
                        + \vec N \leftcontract \vec\del(n) \overeq N 
       + {\cal U}(U,n) \times_n \curl_n \vec N 
      \cr & \qquad \quad 
       - ({\cal U} \leftcontract \{ \Lie(n)\sub{e_0} P(n)^\flat \})^\sharp \cr 
     &\qquad + \vec N \, D\lie(U,n,e_0) \ln \Gamma(U,n) / dt 
     + N^2\gamma(U,n)^{-1} F(U,n) ]^a     \ .\cr
}\end{equation}   

The additional shift spatial covariant derivative term combines with
the existing term in the spatial gravitational force to form twice its
antisymmetric part,  eliminating the contribution of the
symmetric part of the shift tensor gravitomagnetic force term and doubling
the contribution of the gravitomagnetic vector force to yield a term
exactly analogous to the threading point of view expression,
modulo a term quadratic in the shift vector field
(which is the Coriolis term in the case of flat spacetime in rotating
coordinates).
A shift Lie derivative term also adds to the gravitoelectric field to form
an expression analogous to the one in the threading point of view.
The spatial metric Lie derivative term is analogous to the symmetric part
of the threading gravitomagnetic tensor.
An annoying correction factor for the coordinate time parametrization
scales the right hand side in the threading point of view. Multiplying this
out leads to higher order terms in the coordinate velocity appearing in
the threading point of view expression. 

In both cases the logarithmic
gravitoelectric potential $ \Phi(o)=\ln L(o)$ maps onto
the potential $\half L(o)^2 \sim \half(L(o)^2 - 1)$ in the coordinate
time representation of the second order acceleration equation, giving
a linear rather than a logarithmic relationship between the square of
the lapse and the potential. Both of these are the same in the
Newtonian limit and agree with the more commonly used reference
splitting definition of M\o ller \cite{mol}
\begin{equation}
       \Phi\refe = \half ( -\four g_{00} - 1 )
\end{equation}   
but differ at post-Newtonian order.

The changes to the slicing point of view acceleration equation which occur
when switching to the second-order form
are not surprising since
its expression must result from a transformation of the threading 
point of view expressions, and in the quasi-orthogonal limit the lapse and
shift of the two points of view coincide. Thorne et al \cite{thoprimac}
discuss these changes
in the slicing point of view second-order acceleration equation for
black hole spacetimes in the weak field slow motion limit.
Forward \cite{for}
briefly discusses a reference decomposition of the second-order
acceleration equation before linearizing to go to the same limit for
an isolated body.

The transformation law relating the gravitoelectric and gravitomagnetic
vector fields in the two points of view may be evaluated either directly
from their definitions, re-expressing each in terms of the other
point of view, or by applying the general congruence point of view
transformation law for the
relative observer boost between $n$ and $m$, making the nonhomogeneous
terms explicit. The result is
\begin{equation}\eqalign{ 
     \vec g(n) 
    &= \gamma{}^2 P(n,m) \{ \vec g(m) + [ \Lie(m)\sub{e_0} \overeq M ]^\sharp
         + \half \nu(n,m) \times_m \vec H(m)  \cr &\quad
         + \half \nu(n,m) \leftcontract M \SYM \vec\del(m) \overeq M 
      \cr & \quad  
    - \gamma{}^{-2} M^{-1} \nu(n,m) \Lie(m)\sub{e_0} \ln (M\gamma) \}\ ,\cr
    \half \vec H(n,e_0)
     &=  P(m,n)^{-1} \{  \half \vec H(m) 
     - \nu(n,m) \times_m \left[ \vec g(m) 
           + \half [ \Lie(m)\sub{e_0} \overeq M ]^\sharp     \right] \} \ , \cr
}\end{equation}   
where $\gamma = \gamma(n,m) = N/M$,
or the inverse transformation
\begin{equation}\eqalign{ 
   \vec g(m)
    &= \gamma{}^2 P(n,m)^{-1}  \{
          \vec g(n) 
           - \gamma{}^2 N^{-2} [ \Lie(n)\sub{e_0} \overeq N ]^\sharp
           + \half \nu(m,n) \times_n \vec H(n)  \cr &\quad
           +  \nu(m,n)\leftcontract N^{-1} \SYM \vec\del(n) \overeq N
      \cr & \quad  
     - \gamma {}^{-2} N^{-1} \nu(m,n) 
            \Lie(n)\sub{e_0} \ln ( \gamma N^{-1})  \}
         \ ,\cr
   \half \vec H(m) &= \gamma{}^2 P(n,m)^{-1} \{ \half \vec H(n,e_0) 
     - \nu(m,n) \times_n \left[ \vec g(n) 
           -\half N^{-1} [\Lie(n)\sub{e_0} \, N^{-1} \overeq N ]^\sharp
      \right.\cr
       &\qquad     +  \left.
             [\nu(m,n) \leftcontract \SYM N^{-1} \del(n) \overeq N ]^\sharp
               \right] \} \ .
}\end{equation}   

\section{Stationary spacetimes and Fermat's principle}

Stationary spacetimes admit a timelike Killing vector field on some open
submanifold. Choosing the threading vector field $e_0$ to be such a Killing
vector field leads to a ``stationary" parametrized nonlinear reference frame
whose parametrization is adapted to the flow of this vector field.
The threading point of view is valid everywhere that it is timelike.

In the computational frame, all stationary fields will have components which
are independent of $t$, and the spatial fields which result
from the measurement of a stationary spacetime field reduce to time-independent
fields on the computational 3-space. 
Spatial differential operators reduce
to the obvious time-independent operators there as well, with
$\Lie(m)\sub{e_0}$ reducing to the ordinary time (parameter) derivative
$d/dt$.
In other words as described by Gerosh \cite{ger},
the algebra of stationary spatial fields is isomorphic
to the tensor algebra on the computational 3-space with the Riemannian geometry
of the time-independent projected spatial metric, expressable as
$\gamma_{ab} dx^a \otimes dx^b$ in local adapted coordinates identified with
their projections down to the computational 3-space. The spatial operators
$\grad_m,\curl_m,\div_m$ of stationary
spatial fields reduce to the corresponding
operators defined with respect to this metric ($\grad(\gamma),\curl(\gamma),
\div(\gamma)$ defined in terms of its connection $\del(\gamma)$)
when projected down to the computational 3-space, as introduced by
Landau and Lifshitz \cite{lanlif41,lanlif75}.

The expansion tensor
\begin{equation}
  \theta(m)_{\alpha\beta} 
   =  \half M^{-1} [ P(m)\Lie\sub{e_0} \four g]_{\alpha\beta} =0
\end{equation}   
vanishes, leading to the agreement of the two temporal operators
$\del\cfw(m)$ and $\del\lie(m)$ when acting on spatial fields,
while the acceleration admits a potential in the ordinary sense
\begin{equation}
  a(m)^\flat = d(m) \ln M = d \ln M \ .
\end{equation}   
The vorticity vector field
\begin{equation}
      \vec \omega(m) = \half M \curl_m \vec M
\end{equation}   
reduces to the ordinary curl of the shift vector field when projected
down to the computational 3-space, corrected by the lapse function.
Thus the gravitoelectric and gravitomagnetic
fields in the threading point of view admit scalar and vector potentials
on the computational 3-space in the ordinary sense. In the static case
where the vorticity vanishes, one may choose the slicing so that the shift
is zero, leading to a static nonlinear reference frame.
If the slicing is also spacelike, then the slicing point of view holds and
one can repeat the discussion for the corresponding quantities with some
differences.

For both points of view
introduce the conformally rescaled spatial metric,
or ``optical metric"
\begin{equation}\eqalign{
    \tilde P(o)^\flat &= L(o)^{-2} P(o)^\flat \ ,\cr
    \tilde \gamma_{ab} &= M^{-2} \gamma_{ab}\ ,\quad
    \tilde g_{ab} = N^{-2} g_{ab} \ .\cr
}\end{equation}   
This re-definition of the spatial metric variable
makes the square of the lapse function an overall conformal
factor of the spacetime metric, which is 
relevant to the conformally invariant
null geodesic problem for stationary spacetimes.

M\o ller \cite{mol} has shown that the coordinate light travel time may 
be used as
a parametrization independent action integral for this problem, giving
a general relativistic generalization of Fermat's principle.
Since the differential of spacetime arclength vanishes along a null geodesic
\begin{equation}
   -\four d s^2 = d\tau\subpmath{o}{}^2 
          - d \ell\subpmath{o}{}^2 = 0 \ ,
\end{equation}   
using an obvious notation for its splitting 
in each point of view, one may solve this for $dt$
(choosing the future-directed root), leading to the
action integral between two fixed points of the computational 3-space.
For the threading point of view one finds
\begin{equation}\label{eq:nullactionsqrt}
  \Delta t = {\textstyle\int} \, dt 
      = {\textstyle\int} [ M^{-1} d\ell\subpmath{m} + M_a dx^a ]
\end{equation}   
or equivalently
\begin{equation}\label{eq:nullactionsqrtpar}
  \Delta t = {\textstyle\int} n\refr{}\subpmath{m} d\ell\subpmath{m}
           = {\textstyle\int} \tilde n\refr{}\subpmath{m} 
                     d\tilde\ell\subpmath{m} \ ,
\end{equation}   
where
\begin{equation}\eqalign{
      n\refr{}\subpmath{m} &= M^{-1} ( 1 + M M_a dx^a/ d\ell_m ) \ ,\cr
     \tilde n\refr{}\subpmath{m} &= ( 1 + M_a dx^a/ d \tilde\ell_m ) \ ,\cr
}\end{equation}   
and $\ell\subpmath{m}$ 
and $\tilde\ell\subpmath{m}$ are the spatial arclength parameters
with respect to the threading spatial metric and optical spatial metric
respectively on the computational 3-space.
Re-expressing these same quantities in terms of the slicing variables leads
to more complicated expressions
\begin{equation}
  \Delta t = {\textstyle\int} n\refr{}\subpmath{n} d\ell\subpmath{n} 
        = {\textstyle\int} \tilde n\refr{}\subpmath{n} 
              d\tilde\ell\subpmath{n} \ ,
\end{equation}   
where
\begin{equation}\eqalign{ 
      n\refr{}\subpmath{n}
            &= N^{-1}[1-N^{-2} N_b N^b]^{-1} 
   \{ [ 1 - N^{-2} N_b N^b 
      \cr & \quad  
 + (N^{-1}N_a dx^a/d\ell\subpmath{n})^2 ]^{1/2}
        + N_a dx^a/ d\ell\subpmath{n} \} \ ,\cr
     \tilde n\refr{}\subpmath{n} &=  [1-\tilde g{}^{ab} N_a N_b]^{-1}
  \{ [ 1 - \tilde g^{ab} N_a N_b + (N_a dx^a/d\tilde\ell\subpmath{n})^2 ]^{1/2}
      \cr & \quad  
        + N_a dx^a/ d\tilde\ell\subpmath{n} \} \ .\cr
}\end{equation}   

These may be interpreted in two ways. For example, in the static case
the action integral reduces to the arclength of the curve in the
computational 3-space with respect to the optical metric, i.e., null geodesics
project down to geodesics of the optical geometry, using the terminology
recently introduced by Abramowicz et al 
\cite{abrcarlas}--\cite{pracha}.
Perlick has introduced the alternate name ``Fermat metric" for the
optical metric \cite{per89,per90a,per90b}.
Alternatively one may interpret the
lapse as introducing an effective index of refraction
as discussed by M\o ller \cite{mol}.
In the nonstatic case this index of refraction becomes anisotropic,
deflecting the paths of light rays from the geodesics of the optical
geometry.

For the static case using a static parametrized nonlinear reference frame,
one can also re-express the acceleration equation in terms of the
optical geometry. If one 
introduces  the ``coordinate momentum" (per unit mass) components
$\tilde{\cal P}(U,m)^a = \tilde {\cal E}(U,m) \dot x{}^a$, then the canonical
momenta are obtained by lowering the index with the optical metric
$\pi_a = \tilde\gamma_{ab} \tilde{\cal P}(U,m)^b$.
The coordinate energy $\tilde{\cal E}(U,m)$ is conserved in the stationary case
since the Lagrangian is independent of $t$, and it is related to the
coordinate momentum by
\begin{equation}
   \tilde{\cal E}(U,m)^2 
    = M^2 \tilde m{}^2 
          + \tilde\gamma_{ab} \tilde{\cal P}(U,m)^a \tilde{\cal P}(U,m)^b \ ,
\end{equation}   
where $\tilde m$ is the ``mass per unit mass", i.e., 1 for a timelike
curve and 0 for a null curve, for which this relation may be used to
define the coordinate energy.

Introduce also the corresponding optical derivative using the optical
spatial connection $\tilde\del(\gamma)$ instead of the spatial metric 
connection $\del(\gamma)$ on the computational 3-space.
Then the static case second order acceleration equation takes the form
\begin{equation}\eqalign{
   \tilde D\lie(U,m)^2 x^a / dt^2
&=
     \tilde{\cal E}(U,m)^{-1} \tilde D\lie(U,m) 
       [ \tilde{\cal E}(U,m) \dot x{}^a ] / dt \cr
&= -(\tilde m/ \tilde {\cal E}(U,m))^2  \, 
                    \tilde\del(m)^a \half[ M^2-1] \cr
     &\qquad + M/{\cal E}(U,m) \tilde\gamma{}^{ab} F(U,m)_b  \ .\cr
}\end{equation}   
For zero rest mass $\tilde m=0$ and no applied force,
this reduces to the geodesic equation for the optical geometry with
the coordinate time as an affine parameter, describing null geodesics as
discussed in exercise 40.3 of Misner, Thorne and Wheeler \cite{misthowhe}.
The above null geodesic action integral 
is just the optical arclength function for this static case,
first studied by Weyl \cite{wey}.

In fact the general
action integral (\ref{eq:nullactionsqrt}) is parametrization independent.
If $\lambda$ denotes a parameter, and $\dot f= df/d\lambda$, then
this action is
\begin{equation}\eqalign{
    {\textstyle\int} \, [ d \tilde\ell\subpmath{m} + \overeq M ]
 &=       {\textstyle\int} \, 
           [ (\tilde\gamma_{ab}\dot x{}^a\dot x{}^b)^{1/2}
                 + M_a \dot x{}^a ] \, d\lambda \ .\cr
}\end{equation}   
For the optical arclength parametrization
$ d \tilde\ell\subpmath{m} / d\lambda =1$ or 
$ \tilde\gamma_{ab} \dot x{}^a \dot x{}^b = 1$, one can use instead the
equivalent action
\begin{equation}\eqalign{
 & {\textstyle\int} \, [ \half \tilde\gamma_{ab}\dot x{}^a\dot x{}^b
                 + M_a \dot x{}^a ] \, d\lambda \ .\cr
}\end{equation}   
Perlick \cite{per90a}
has noted that each of these actions continue to be valid in the
nonstationary case, 
but then of course cannot be considered without the  evolution equation for 
$t$ as well. Clearly since the null geodesics are conformally invariant,
it is enough to have a conformally stationary spacetime for the problem
to reduce to a purely spatial one, allowing the same analysis to extend
to interesting cosmological spacetimes \cite{per90b}.
One can repeat the discussion for the force equation
for timelike test particles for the case of lightlike test particles and
obtain the general second-order equation and the spatial force equation
for that case and also develop the Lagrangian approach to the former case.
This has not been done here for reasons of space.

Note that the shift plays a role similar to the effect of the
vector potential in electromagnetism.
Samuel and Iyer \cite{samiye} and Perlick \cite{per90b,per91}
have explored the analogy between the gravitomagnetic field and the
magnetic field for stationary spacetimes.

\section{Post-Newtonian approximation}

The post-Newtonian treatment of weak gravitational fields within
general relativity and its parame\-trized post-Newtonian (PPN) generalization
are based on a preferred class of local coordinate systems defined
by certain functional conditions on the metric components.
These preferred coordinates $\{x^\alpha\} = \{t,x^a\}$ naturally
introduce the structure of a ``post-Newtonian" nonlinear reference frame
and a set of gauge transformations among different choices of such frames.
Since these nonlinear reference frames are ``quasi-orthogonal,"
both the slicing and threading points of view are valid and the
slicing and threading variables are closely related.

Assuming the notation of Misner, Thorne and Wheeler \cite{misthowhe},
the usual post-Newtonian conditions on the coordinate components of 
the metric in coordinates adapted to a post-Newtonian nonlinear
reference frame are the following, using the abbreviated notation
$O(n)\equiv O(\epsilon^n)$
\begin{equation}\eqalign{
   \four g_{00} &= - 1 - 2 \Phi + O(4) = - M^2 \ ,\cr
   \four g_{0a} &= \Phi_a + O(5)       = N_a \ ,\cr
   \four g_{ab} &= \delta_{ab} + O(2)  = g_{ab} \ ,\cr
}\end{equation}   
and
\begin{equation}\eqalign{
   \four g^{00} &= - 1 + 2 \Phi + O(4) = - N^{-2} \ ,\cr
   \four g^{0a} &= \delta^{ab}\Phi_b + O(5)       = M^a \ ,\cr
   \four g^{ab} &= \delta^{ab} + O(2)  = \gamma^{ab} \ ,\cr
}\end{equation}   
where $\Phi \sim O(2)$ and $\Phi_a \sim O(3)$. 
The metric components $(\four g_{00}, \four g_{0a},\four g_{ab})$ are
respectively (even,odd,even) in order and are cut off at orders
$(O(4),O(5),O(2))$; the same is true respectively of the lapse, shift and
spatial metric components in each point of view.
This leaves only the $O(4)$ term in $\four g_{00}$
and the order $O(2)$ terms in $\four g_{ab}$ to make explicit.

Because of the relationships
\begin{equation}\eqalign{
      N &= M [ 1- M^2 M_a M_b\gamma^{ab} ]^{-1/2} \ ,\cr
      N_a &= M^2 M_a \ ,\cr
      g_{ab} &= \gamma_{ab} + M^2 M_a M_b \ ,\cr
}\end{equation}   
the lapses and the spatial metric components 
(even order) agree through $O(4)$, while the shift components agree
through order $O(3)$, both with each other and with the obvious corresponding
variables defined using the reference decomposition
of either the spacetime metric or inverse metric. In other words
the slicing and threading metric variables
agree up to the first post-Newtonian order and effectively
reduce to the corresponding reference variables.
The projected computational spatial frame vectors and 1-forms in the two
points of view differ from the computational ones by 
\begin{equation}\eqalign{
     \epsilon_a - e_a & \sim O(3) \partial_0 \ ,\qquad
      \theta^a - \omega^a   \sim O(3) dt \cr
}\end{equation}   
so the distinction between them is lost in the shift and spatial metric fields.
The relative velocity satisfies $||\nu(m,n)|| \sim O(3)$, leading to the
quasi-orthogonal condition on the nonlinear reference frame and the agreement
of the slicing and threading metric variables.

Blanchet and  Damour \cite{bladam}
fix the definition of the potential $\Phi$ to $O(4)$ by the
condition that it coincide with the fully nonlinear
gravitational potential
\begin{equation}
   \Phi = \ln L(o) + O(6)  \ ,
\end{equation}   
or
\FL
\begin{equation}\eqalign{
    L(o) &= e^{\Phi} + O(6) 
         = 1 + \Phi + \half \Phi^2 + O(6) \ , \cr
}\end{equation}   
which is the same to the first post-Newtonian order in both points of view,
so that
\FL
\begin{equation}
    -\four g_{00} = e^{2\Phi} + O(6) 
         = 1 + 2\Phi + 2 \Phi^2 + O(6) \ .
\end{equation}   
The gravitoelectric and gravitomagnetic fields then have the following
behavior
\begin{equation}\eqalign{
     H(m)^a &= H^a + O(5) 
                 = H(n,e_0)^a + O(5)\ ,\cr
     g(m)_a &= g_a + O(6) \ ,\cr
     g(n)_a &= -\partial_a \Phi + O(6) \ ,\cr
}\end{equation}   
where the lowest order post-Newtonian threading fields are defined by
\begin{equation}\eqalign{
      H^a &= \epsilon^{abc} \partial_b \Phi_c  \ ,\quad 
      g_a = -\partial_a \Phi - \partial_0 \Phi_a \ ,\cr
}\end{equation}   
The gravitomagnetic fields agree to first post-Newtonian order but
the gravitoelectric fields differ by an $O(4)$ time derivative term.
For some reason independent of which point of view people favor for the
full Einstein equations, it is the post-Newtonian threading fields which
are always used without comment \cite{bracavtho,damsofxua,bladamsch},
and which agree with Forward's reference decomposition of the 
gravitational variables \cite{for}.

In the post-Newtonian approximation the 4-potential for the
threading gravitoelectromagnetic vector fields is just
\FL
\begin{equation}\eqalign{
&       m^\flat  =- M (dt - M_a dx^a) \cr&\quad
           \to - dt + [ ( -\Phi + O(4) ) dt + (\Phi_a + O(5) dx^a ]  \ .\cr
}\end{equation}   
The explicit  terms in the square bracketed expression
define the post-Newtonian 4-potential introduced by
Damour et al \cite{damsofxua}.

The threading spatial gauge freedom to change the slicing reduces to the
usual electromagnetic gauge transformations of the scalar and vector
potentials $(\Phi,\Phi_a)$
\begin{equation}\eqalign{
     t &\mapsto t + \Lambda  + O(5)\ , \quad \Lambda \sim O(3) \ ,\cr
     \Phi &\mapsto \Phi + \partial_0 \Lambda  + O(6) \ ,\cr
     \Phi_a &\mapsto \Phi_a + \partial_a \Lambda + O(5) \ .\cr
}\end{equation}   
This leaves the threading gravitoelectric and gravitomagnetic fields invariant.

Of the kinematical fields, only the expansion tensor remains 
to be evaluated and its
form depends on the spatial metric. It is
of order $O(3)$, differing in the two points of view at that
order by spatial derivatives of the shift. To get a handle on it,
one must examine the post-Newtonian restrictions on the spatial metric.

Introduce the ``anti-optical" spatial metric in both points of view by
\begin{equation}\eqalign{
    \tilde P(o)^\flat &= L(o)^{2} P(o)^\flat \ ,\cr
    \tilde \gamma_{ab} &= M^{2} \gamma_{ab}\ ,\quad
    \tilde g_{ab} = N^{2} g_{ab} \ .\cr
}\end{equation}   
To post-Newtonian order the distinction between slicing and threading
is unimportant for the spatial metric.
The threading ``anti-optical" spatial metric is
associated with the generalized Lewis-Papapetrou form of the spacetime
metric 
\cite{lew}--\cite{kraetal} 
advocated by Perj\'es \cite{per};
it naturally arises in the analysis of the initial value problem in
that point of view, which is the problem to which the Einstein
equations reduce in the stationary case \cite{kraetal}.
Blanchet and Damour \cite{bladam,bladamsch}
have noticed that the threading anti-optical metric
plays a privileged role in the post-Newtonian analysis.
Damour et al \cite{damsofxua,damsofxub} explicitly define this metric by the
relation
\begin{equation}
    \tilde\gamma^{1/2} \tilde\gamma^{ab} = 
         M \gamma^{1/2} \gamma^{ab} = \four g^{1/2} \four g^{ab}
\end{equation}   
satisfied by the threading anti-optical metric, and they
explain its importance using the simple expression for the Einstein
tensor written in terms of $\four g^{1/2} \four g^{\alpha\beta}$.

This may also be seen directly from the slicing/threading expression
for the Einstein tensor. Since the projected computational frame is
an observer-adapted frame, setting $u=o$ in \Eq (\ref{eq:einstein})
provides the relevant formulas.
To post-Newtonian order, all of the spatial curvature tensors agree
with the slicing one which is the usual curvature of the induced metric
on the slice,  since the differences are of order $O(6)$ \cite{jancar91}.
The lowest order terms in the spatial projection of the spacetime
Einstein tensor in the slicing point of view are 
\FL
\begin{equation}
    G\sym(n)^a{}_b - \del(n)_b a(n)^a + \delta^a{}_b \del(n)_c a(n)^c
         \sim O(2) \ .
\end{equation}   
Under the  conformal transformation which defines the slicing anti-optical
metric,
the Einstein tensor has the following transformation law \cite{kraetal}
\begin{equation}\eqalign{
    \tilde G(n)_{ab} &= G(n)_{ab} - [ \del(n)_a - a(n)_a ] a(n)_b 
      + g_{ab} \del(n)^c a(n)_c   \ .\cr
}\end{equation}   
Thus to the lowest post-Newtonian order $O(2)$ (neglecting the terms
quadratic in $a(n)$),
the spatial projection of the spacetime Einstein tensor is just
the Einstein tensor of the anti-optical metric, which must vanish
to that order since the spatial projection of the energy-momentum
tensor is of order $O(4)$.
The anti-optical metric (slicing or threading) 
is therefore flat to order $O(2)$.
This is the key observation of Damour et al \cite{damsofxua},
who choose the obvious gauge condition that the spatial coordinates
be Cartesian with respect to the anti-optical metric to that order
\begin{equation}
    \tilde\gamma_{ab} = \delta_{ab} + O(4)
\end{equation}   
or equivalently
\begin{equation}
    \gamma_{ab} = M^{-2}  \delta_{ab} + O(4) 
     = [1-2\Phi] \delta_{ab}  + O(4)\ .
\end{equation}   

The operations
$\cdot$ , $\times$  and $\vec\del$ will denote the flat space operations
in these special coordinates. The post-Newtonian threading gravito-vector
fields then have the definition
\begin{equation}\eqalign{
        \vec H &= \vec\del \times \vec\Phi \ ,\quad 
        \vec g = - \vec\del \Phi - \partial_0 \vec\Phi \ ,\cr
}\end{equation}   
which have the consequences
\begin{equation}\eqalign{
     \vec\del \cdot \vec H = 0 
         = \vec\del \times \vec g + \partial_0 \vec H \ .
}\end{equation}   
These are just the post-Newtonian limit of \Eqs (\ref{eq:gemeq}).

The expansion tensor then has the behavior
\begin{equation}\eqalign{
    \theta(m)_{ab} &= - \partial_0 \Phi \, \delta_{ab} + O(5) \ ,\cr
    \Theta(m)      &= - 3 \partial_0 \Phi + O(5) \ .\cr
    \theta(n)_{ab} &= - \partial_0 \Phi \, \delta_{ab} 
                       - \del_{(a} \Phi_{b)}  + O(5) \ ,\cr
    \Theta(n)      &= - 3 \partial_0 \Phi 
            - \vec \del \cdot \vec \Phi + O(5) \ .\cr
}\end{equation}   

The remaining
Einstein equations under these conditions resemble the remaining
half of Maxwell's equations.
The Ricci form of the Einstein equations with the trace-reversal
of the energy-momentum tensor on the right-hand side is more convenient
to obtain these equations
\FL
\begin{equation}
        \four R^{\alpha\beta} 
     = 8\pi [ \four T\TR]^{\alpha\beta} 
     = 8\pi [ \four T^{\alpha\beta} 
       - \half g^{\alpha\beta} \four T^\gamma{}_\gamma] \ .
\end{equation}   
The remaining two linearly independent Einstein equations
in the post-Newtonian approximation expressed in terms
of the projected computational frame
\FL
\begin{eqnarray}
     \four R^{\top\top} 
         &=& - \vec\del \cdot \vec g + 3 \partial_0{}^2 \Phi + O(6) 
          = 8\pi [\four T\TR]^{\top\top}\ ,\nn\\
     2\four R^\top{}_a
         &=& [- \vec\del \times \vec H + 4 \partial_0 \vec g ]_a + O(5) 
          = 16\pi [\four T\TR]^\top{}_a \ .
\end{eqnarray} 
are  the second pair of Maxwell-like equations for the gravitoelectromagnetic
vector fields given by
Braginsky et al \cite{bracavtho} and
Damour et al \cite{damsofxua}.

The ``standard post-Newtonian gauge condition" for the time coordinate $t$
comes from identifying a time derivative in the $O(4)$ behavior of
$\four R^{\top\top}$
\begin{equation}
      \four R^{\top\top} =  \del^2 \Phi + \partial_0 
            [ \vec\del \cdot \vec \Phi + 3 \partial_0 \Phi + O(5)]  \ .
\end{equation}   
Setting the expression in square brackets to zero gives this condition,
and its imposition
leads to a simple Poisson equation for the scalar potential.

The harmonic gauge condition for $t$
has the following representations
\begin{equation}\eqalign{
    0 &= \four \del_\alpha \del^\alpha t 
          = \partial_\alpha [ \four g^{1/2} \four g^{0\alpha} ]
      =\left\{\eqalign{
        &  M \gamma^{1/2} [ \div_m \vec M - \vec g(m) \cdot_m \vec M
              \cr&\quad 
            - M^{-1} \Lie(m)\sub{m} \ln ( \gamma^{1/2} M^{-1} ) ] \cr
       & N^{-1} g^{1/2} [ \div_n \vec N - \vec g(n) \cdot_n \vec N
              \cr&\quad 
            - N^{-2} \Lie(n)\sub{e_0} \ln ( g^{1/2} N^{-1} ) ] \ ,\cr}
        \right.
}\end{equation}   
with the post-Newtonian limit
\begin{equation}
        \partial_0 \Phi + \vec\del \cdot \vec \Phi + O(5) = 0 \ .
\end{equation}   
This differs only by a numerical factor from the standard gauge condition.

\section{Schiff precession formula}

The discussion of gyroscope precession presented above
is valid for any spacetime and expresses the
angular velocity of the
relative rotation of the spin vector with respect to the observer congruence
or subsequently with respect to a preferred orthonormal observer-adapted
frame whose spatial part undergoes co-rotating Fermi-Walker
transport along the observer congruence. The spatial distribution of the
orientation of such a spatial orthonormal frame is still arbitrary.
For the first result to be meaningful one must have a preferred observer
congruence, and for the second, a preferred distribution of the orientation
of the spatial orthonormal frame.

Stationary spacetimes have a preferred observer congruence associated
with any timelike Killing vector field, leading to a stationary observer
congruence with a stationary 4-velocity $u$.
Since spatial Lie transport along such a congruence
coincides with co-rotating Fermi-Walker
transport, the spatial projection of any frame which is comoving with respect
to $u$ will yield a spatial frame which
is spatially comoving and which undergoes co-rotating Fermi-Walker
transport along $u$. In particular comoving coordinates whose spatial
coordinates are orthogonal yield such a frame under spatial projection
which can be normalized to an observer-adapted orthonormal frame with
the same properties.

For a stationary spacetime
representing an isolated mass distribution
which is asymptotically flat at spatial infinity
one can pick out a preferred Killing vector (in the event of additional
symmetry),
namely the one which reduces to the unit vorticity-free
timelike Killing vector of the
asymptotic geometry with respect to which the isolated body is  not moving.
This leads to the static observer congruence.
The choice of a spatial orthonormal frame is less clear. A ``Cartesian-like"
frame would be preferable but no canonical choice exists.
In the post-Newtonian theory or its parametrized generalization,
one works with
a class of ``Cartesian-like" coordinates involving a gauge freedom,
so such a frame is available, modulo these gauge transformations.
Nester has recently shown that a preferred slicing orthonormal frame
exists which asymptotically approaches a given spatial Cartesian frame
on an asymptotically flat slice \cite{nes89,nes91}.
Its boost could be taken to define a preferred 
``Cartesian-like" threading orthonormal frame 
if one has a preferred slicing within the class of asymptotically flat
nonlinear reference frames.

For black hole spacetimes, and indeed the wider class of stationary axially
symmetric spacetimes,
a preferred class of stationary orthonormal
spatial  frames does exist in both the slicing and threading points of view.
Consider a black hole spacetime in the Boyer-Lindquist coordinate system,
with its associated nonlinear reference frame.
The threading point of view holds outside the ergosphere where the Killing
observers follow the timelike time lines, while the slicing point of view
holds outside of the event horizon where the slicing is spacelike.
The stationary
threading observers have the interpretation of being nonrotating with
respect to the asymptotically flat region of spacetime and are called
the static observers,
while the nonstationary
slicing observers have the interpretation of being locally nonrotating
with respect to the spacetime geometry.

The Boyer-Lindquist spatial coordinates $\{r,\theta,\phi\}$ are orthogonal
so both the coordinate derivatives $\{e_a\}$ and coordinate differentials
$\{\omega^a\}$ are orthogonal and can be normalized and then completed
uniquely to an (axially symmetric stationary)
orthonormal spacetime frame or dual frame. Normalizing the spatial
coordinate derivatives leads
to the slicing orthonormal frame $\{n,e_{\hat a}\}$ 
with dual frame $\{\omega^\bot,\theta^{\hat a}\}$ 
while normalizing the spatial coordinate differentials leads to the
threading orthonormal frame $\{m,\epsilon_{\hat a}\}$
with dual frame $\{\omega^\top, \omega^{\hat a}\}$.

One can boost each
of these two orthonormal frames uniquely to align them with the 4-velocity
of an arbitrary gyro worldline
\FL
\begin{equation}\eqalign{
  B(u,n) \{ n,e_{\hat a} \} &= \{ u,E_{({\rm sl})a} \}\ ,\cr
  B(u,m) \{ m,\epsilon_{\hat a} \} 
    &= \{ u,E_{({\rm th})a} \}
  = B(u,m) B(m,n) \{ n,e_{\hat a} \} \ .\cr
}\end{equation}   
The two orthonormal frames so obtained are related to each other
by the time-dependent Thomas rotation determined by the composition
of the two boosts $B(u,m)$ and $B(m,n)$
\begin{equation}\eqalign{
     E_{({\rm th})a} &= B(u,m) B(m,n) B(n,u) E_{({\rm sl})a}
               = R(u,m,n) E_{({\rm sl})a} \ , \cr
}\end{equation}     
which may in some sense be interpreted as the relative rotation of
the spatial axes of the slicing and threading observers as determined by
the gyro.
The boosted frame in each point of view is the spatial
frame that an observer following the worldline of the gyro would
reconstruct as the frame he would see if 
that frame were not moving relative to him.

In the slicing point of view, 
the co-rotating Fermi-Walker relative angular velocity
measures the
precession of the spin relative to the locally nonrotating observers,
while in the threading point of view, it is instead relative
to the static
Killing observers which in some sense reflect the properties
of the nonrotating frame of the ``distant stars"
(whose incoming light rays have fixed direction with respect to a
co-rotating Fermi-Walker transported frame along these observers'
worldlines).
However, the spatial frames described above
are ``spherical" in nature rather than
Cartesian so the space curvature precession also includes the rotation
of the observer frame relative to Cartesian-like frames along the gyro
worldline.

All asymptotically flat axially symmetric stationary spacetimes have 
such a preferred stationary nonlinear reference frame whose slicing is
orthogonal to the locally nonrotating observers and whose threading
is along the static observers \cite{greschvis}.
A similar situation exists in the PPN theory, where the PPN spatial
coordinates are orthogonal to the lowest nontrivial order as well as
Cartesian-like, so one can introduce a preferred class of
slicing and threading orthonormal frames. The orthonormal threading frame
is given in section 39.10 of Misner, Thorne and Wheeler \cite{misthowhe}.

The classic spin precession formula of Schiff \cite{sch}
describes how the spin vector precesses relative to the ``distant stars"
as seen by an geodesic
observer carrying the gyro in the gravitational field of an
isolated body.
It may be obtained by evaluating in the post-Newtonian order
the formula (\ref{eq:frameangvel}) for the angular
velocity $\zeta(U,u,e)$ relative to the threading orthonormal 
frame.
One only need evaluate the space curvature precession term to this order.

The key difference with the limiting expression for $\zeta\cfw$ is the
fact that in this limit within general relativity,
the space curvature precession has twice the
value of the spin orbit precession, leading to a total coefficient of
$\fraction32$.
It has been generalized to the PPN theory as discussed by Misner, Thorne
and Wheeler \cite{misthowhe} or Weinberg \cite{wei}.

To post-Newtonian order, the slicing
spatial orthonormal frame is
$ e_a = (1+ \Phi) \partial/\partial x^a$, so the
spatial structure functions,
the spatial connection components and the space curvature
precession in that point of view are
\begin{equation}\eqalign{
   C^a{}_{bc} &=  2\delta^a{}_{[b} g(n)_{c]} + O(4) \ ,\cr
   \Gamma(n)_{abc} &=  2\delta_{b[a} g(n)_{c]} + O(4) \ ,\cr
    \zeta\sc(U,n,e){}^a &= \eta(n)^{abc} \nu(U,n)_b g(n)_c + O(5) \cr
      &= [ \nu(U,n) \times_n \vec g(n) ]^a + O(5) \cr
      &= [ \nu(U,m) \times_m \vec g(m) ]^a + O(5)\ .\cr
}\end{equation}   

Thus in this limit the space curvature precession, which has a
completely different
origin from the spin-orbit precession, has twice the magnitude of the latter
precession, leading to the total factor $\fraction32$
\FL
\begin{eqnarray}
   \zeta\gyro(U,m,e) & \rightarrow &
  -\half \vec H(m)
   -\half \nu(U,m) \times_m  a\cfw(U,m) 
        \nn\\ &&\quad 
       + \fraction32 \nu(U,m) \times_m \vec g(m)  \ .
\end{eqnarray}

The actual Schiff formula is obtained by substituting explicit expressions
for the post-Newtonian gravitoelectromagnetic potentials.
Its verification is the goal 
of the long awaited Stanford gyroscopic precession experiment \cite{eve}
and of the proposed LAGEOS experiment \cite{ciu}.
These and other experiments \cite{brapoltho,nor,maspaiwil}
have provided much of the motivation for talking
about ``gravitomagnetism."

\section{Conclusions}

A single framework has been introduced which encompasses all possible
approaches to space-plus-time splittings of spacetime and allows
transformations between them to be considered.
Precise relative observer maps and differential operators have been
introduced which first determine the definition of spatial gravitational
forces and then neatly characterize the gyro precession formula
in terms of them.
In the post-Newtonian approximation, all of
these various spatial gravitational forces are closely related, but it is the
threading forces which are universally used in the
application of that approximation.
By examining the origin of the post-Newtonian equations in the fully nonlinear
context of a parametrized nonlinear reference frame, a better understanding
of their structure is obtained.

This same scheme can be used in studying 
the Sagnac effect \cite{pos,ashmag}
and the closely related synchronization gap \cite{hennel},
Maxwell's equations for the electromagnetic 
field \cite{ell73,ben,thomac,zha},
and the fully nonlinear Einstein equations and their initial value problem. 
The initial value problem for the
threading point of view is still not well understood \cite{ferr89,sta},
although it is closely related to the exact solutions work for
stationary spacetimes \cite{kraetal}.
The perturbation problem for Friedmann-Robertson-Walker models
has been considered both in the slicing \cite{durstr,bar80}
and the more general congruence \cite{haw,ellbru,ellhwabru}
points of view; the present formalism allows one to more easily relate the
two. Similarly the idea of a Newtonian limit \cite{ehlnew,lot}
crucially relies on a
family of spacetime splittings, for which the present language is
rather helpful in describing. 

Rotation in general relativity has intrigued people for quite some time,
but some
rather simple rotational aspects of familiar exact solutions have still not
been clearly presented. Examination of black hole spacetimes, the G\"odel
spacetime \cite{god}
and Minkowski spacetime in a uniformly rotating nonlinear
reference frame using the present approach leads to a more
intuitive understanding of the familiar properties of
these models and how they compare to each other
in terms of the
individual contributions of gravitoelectric, gravitomagnetic and 
space curvature effects in their various representations.
This will be discussed in a subsequent article.

\section*{Acknowledgments}

Two of the authors (P.C. and D.B.) acknowledge support from the Italian
Consiglio Nazionale delle Ricerche (CNR). 
One (D.B.) acknowledges support from the Department of Mathematics
of the University of Rome and the other (P.C.) 
greatly appreciates the kind hospitality of F. Everitt and the Gravity
Probe-B group at Stanford.
All thank the generous support of R. Ruffini and the International
Center for Relativistic Astrophysics at the University of
Rome.



\section*{Corrections}

This reformatted version contains the following misprint corrections of the original article (to which the page numbers refer) and one reference publication update:

\begin{itemize}

\item
p.~3, Section II, first sentence, remove:
$=- \four\eta^{0123}$.

\item
p.~5, Eq.~(2.10) both lines, after first equal sign:\\ 
projections on both indices inserted.

\item
p.~8, end of phrase preceding Eq.~(3.4), superscript before comma should be $b$. 

\item
p.~16, Eq.~(6.12), put tilde over $E$ on left hand side of equation, first line.

\item
p.~17, Eq.~(6.15), change $(U,u)$ to $(u)$ on all left hand sides.

\item
p.~17, Eq.~(6.18), Line 1, change $F\em(u)$ to $F\em(U,u)$; 
line 3, change $H\tem(U,u)$ to $H\tem(u)$.

\item
p.~19, Eq.~(7.8), fourth line added for $R\fw(u)^{[ab]}{}_{cd}$.

\item
p.~21, Eq.~(8.1), change $\gamma P(U,u)$ to $\gamma^2 P(U,u)$ on right hand side of second equation.

\item
p.~23, Eq.~(9.9), change $\vec\omega(U,u)$ to $\vec\omega(u)$.

\item
p.~28, Eq.~(10.9), Line 1, second equation, left hand side, change $\four g^0$ to $\four g^{00}$.

\item
p.~30, Eq.~(10.14), left hand side subscript before equal sign, change $d\tau_{(n,U)}$ to $d\tau_{(U,n)}$.

\item
p.~31, Eq.~(10.16), Line 2, second equation, change
$$
H\lie(n)_{\alpha\beta} = -  N^{-1} \Lie(n)\sub{e_0} g_{\alpha\beta}
$$
to 
$$
H\lie(n)_{\alpha\beta} = -  N^{-1} \Lie(n)\sub{e_0-\vec N} g_{\alpha\beta}
   = -2 \theta(n)_{\alpha\beta}
$$

\item
p.~31, Eq.~(10.20), change $F\g\tem(U,n,e_0)$ to $\tilde F{}\g\tem(U,n,e_0)$.

\item
p.~34, Eq.~(11.8), change $o$ to $m$ and $n$ respectively in right hand side cases expressions.

\item
p.~35, Eq.~(11.12), Line 2, change $d\tau_{(m,U)}$ to $d\tau_{(U,m)}$.

\item
p.~36, Eq.~(11.16), Line 3, change
$(M\gamma_L)$ to $(M\gamma)$.  

\item
p.~40, End of page, change $(O(4),O(3),O(2))$ to $(O(4),O(5),O(2))$.

\item
p.~42, Eq.~(13.20), Lines 1 and 3, change $\partial_{ab}$ to $\delta_{ab}$.

\item
p.~50, Reference 73, change year from (1991) to (1989).

\end{itemize}

\end{document}